\documentclass[twocolumn]{aastex631}

\defcitealias{C19}{C20}
\defcitealias{Col22}{C22}
\usepackage{makecell}

\shorttitle{AndXIX Abundances}
\shortauthors{Cullinane et al.}
\begin{document}

\title{Elemental Abundances in And XIX From Coadded Spectra}

\correspondingauthor{L. R. Cullinane}
\email{lcullinane@aip.de}

\author[0000-0001-8536-0547]{L. R. Cullinane}
\affiliation{Leibniz-Institut f{\"u}r Astrophysik (AIP), An der Sternwarte 16, D-14482 Potsdam, Germany}
\affiliation{The William H. Miller III Department of Physics \& Astronomy, Bloomberg Center for Physics and Astronomy, Johns Hopkins University, 3400 N. Charles Street, Baltimore, MD 21218, USA}

\author[0000-0003-0394-8377]{Karoline M. Gilbert}
\affiliation{Space Telescope Science Institute, 3700 San Martin Dr., Baltimore, MD 21218, USA}
\affiliation{The William H. Miller III Department of Physics \& Astronomy, Bloomberg Center for Physics and Astronomy, Johns Hopkins University, 3400 N. Charles Street, Baltimore, MD 21218, USA}

\author[0000-0002-9933-9551]{Ivanna Escala}
\altaffiliation{Carnegie-Princeton Fellow}
\affiliation{Department of Astrophysical Sciences, Princeton University, 4 Ivy Lane, Princeton, NJ, 08544 USA}
\affiliation{The Observatories of the Carnegie Institution for Science, 813 Santa Barbara St, Pasadena, CA 91101}

\author[0000-0002-3233-3032]{J. Leigh Wojno}
\affiliation{Max-Planck Institute for Astronomy, Königstuhl 17, D-69117 Heidelberg, Germany}

\author[0000-0001-6196-5162]{Evan N. Kirby}
\affiliation{Department of Physics and Astronomy, University of Notre Dame, 225 Nieuwland Science Hall, Notre Dame, IN 46556, USA}

\author[0000-0002-7438-1059]{Kateryna A. Kvasova}
\affiliation{Department of Physics and Astronomy, University of Notre Dame, 225 Nieuwland Science Hall, Notre Dame, IN 46556, USA}

\author[0000-0002-9599-310X]{Erik Tollerud}
\affiliation{Space Telescope Science Institute, 3700 San Martin Dr., Baltimore, MD 21218, USA}

\author[0000-0002-1693-3265]{Michelle L. M. Collins}
\affiliation{Physics Department, University of Surrey, Guildford GU2 7XH, UK}

\author[0000-0003-0427-8387]{R. Michael Rich}
\affiliation{Department of Physics and Astronomy, UCLA, Los Angeles, CA 90095, USA}



\begin{abstract}
	With a luminosity similar to that of Milky Way dwarf spheroidal (dSph) systems like Sextans, but a spatial extent similar to that of ultradiffuse galaxies (UDGs), Andromeda (And) XIX is an unusual satellite of M31. To investigate the origin of this galaxy, we measure chemical abundances for AndXIX derived from medium-resolution (R$\sim$6000) spectra from Keck II/DEIMOS\@. We coadd 79 red giant branch stars, grouped by photometric metallicity, in order to obtain a sufficiently high signal-to-noise ratio (S/N) to measure 20 [Fe/H] and [$\alpha$/Fe] abundances via spectral synthesis. The latter are the first such measurements for AndXIX\@. The mean metallicity we derive for AndXIX places it $\sim2\sigma$ higher than the present-day stellar mass-metallicity relation for Local Group dwarf galaxies, potentially indicating it has experienced tidal stripping. A loss of gas and associated quenching during such a process, which prevents the extended star formation necessary to produce shallow [$\alpha$/Fe]--[Fe/H] gradients in massive systems, is also consistent with the steeply decreasing [$\alpha$/Fe]--[Fe/H] trend we observe. In combination with the diffuse structure and disturbed kinematic properties of AndXIX, this suggests tidal interactions, rather than galaxy mergers, are strong contenders for its formation. 
\end{abstract}



\section{Introduction} \label{sec:intro}

First discovered in the Pan-Andromeda Archaeological Survey \citep[PAndAS:][]{mcconnachieTrioNewLocal2008}, Andromeda (And) XIX is an unusually diffuse dwarf spheriodal (dSph) satellite located 115~kpc from M31 \citep{connBayesianApproachLocating2012}. With a total luminosity of $7.9\times10^5$~L$_\odot$ and a half-light radius of $r_{\text{half}}=3065^{+935}_{-1065}$~pc \citep{martinPAndASVIEWANDROMEDA2016} it is a factor of 10 more extended than similarly luminous Local Group dwarfs such as Sextans and Carina \citep{mcconnachieObservedPropertiesDwarf2012}. Its size is more comparable to ultradiffuse galaxies \citep[UDGs;][]{vandokkumFORTYSEVENMILKYWAYSIZED2015} in more distant galaxy clusters, but with a central surface brightness of $\mu_{V,0}=29.3\pm0.4$~mag arcsec$^{-2}$ \citep{martinPAndASVIEWANDROMEDA2016} it is orders of magnitude fainter than these systems. The extended Milky Way (MW) satellite Antlia II \citep{torrealbaHiddenGiantDiscovery2019} is one of the only known comparably diffuse systems. 

A number of different formation scenarios have been put forward to explain the observed properties of AndXIX\@. Tidal interactions are a possibility: both tidal shocking -- where a system is impulsively heated and subsequently expands during re-virialisation \citep{amoriscoGiantColdSatellites2019,ogiyaTidalFormationDark2022} -- and tidal stripping -- where a loss of mass also results in re-virialisation and subsequent expansion \citep[e.g.][]{bennetEvidenceUltradiffuseGalaxy2018, jacksonDarkMatterdeficientDwarf2021,ogiyaTidalFormationDark2022}  -- are capable of significantly increasing the effective radius of satellites during pericentric passages around a massive host, particularly in galaxy clusters \citep{carletonFormationUltradiffuseGalaxies2019}. Indeed, the similarly-diffuse Antlia II is thought to have been strongly impacted by tides during a recent close pericentric passage around the MW \citep{jiKinematicsAntliaCrater2021}. 

Another proposed scenario is that strong feedback from bursty star formation dynamically heats the dark (and subsequently baryonic) matter distribution of the galaxy \citep[e.g.][]{dicintioNIHAOXIFormation2017, readDarkMatterHeats2019}, producing a diffuse, low-density system. However, AndXIX lacks the extended star formation necessary for this effect; it formed $\sim$50\% of its stars 13.5~Gyr ago and 90\% of its stars $\gtrsim$10~Gyr ago, with no indication of any star formation in the last 8~Gyr \citep[henceforth referred to as C22]{Col22}.

A third possibility is that AndXIX has been produced via galaxy mergers, with several different merger types as possible explanations. Simulations suggest that relatively high-velocity collisions between gas-rich dwarfs can produce diffuse UDG-like galaxies \citep{silkUltradiffuseGalaxiesDark2019,shinDarkMatterDeficient2020,leeDarkMatterDeficient2021,otakiFrequencyDarkMatter2023}. During these collisions, separation is induced between dark and baryonic matter; the latter of which then experiences shock compression to subsequently form stars. A strong burst of star formation in AndXIX $\sim$10~Gyr ago, during which AndXIX formed $\sim40$\% of it stars \citepalias{Col22} is consistent with the expected post-collision starburst in this scenario. 

In alternative merger-related scenarios, simulations show that several “dry” (i.e.\ gas-poor) mergers between smaller satellites, each quenched by reionization, can build diffuse galaxies by depositing stars in the outskirts of the primary galaxy \citep{reyEDGEOriginScatter2019}. So-called Tidal Dwarf Galaxies (TDGs), formed from overdensities in the debris of gas-rich galaxy mergers which become self-gravitating \citep[e.g.][]{ducIdentificationOldTidal2014,bennetEvidenceUltradiffuseGalaxy2018,ploeckingerTidalDwarfGalaxies2018}, can also be diffuse and UDG-like.

Abundance measurements -- and in particular the abundances of $\alpha$-elements (e.g.\ Mg, Ca, Si, Ti) relative to iron -- can help distinguish between these scenarios. Type II (core-collapse) supernovae (SN) are a significant source of $\alpha$-element production, while the ejecta of Type Ia SN are comparatively much richer in iron; as these two processes have differing timescales, the relative abundances of [$\alpha$/Fe] vs.\ [Fe/H] allow us to trace the enrichment history and star-formation timescales within a galaxy \citep[e.g.,][]{tinsleyStellarLifetimesAbundance1979,gilmoreChemicalEvolutionBursts1991}, and identify the remnants of accreted systems \citep[e.g.][]{fontCosmologicalSimulationsFormation2011,naiduEvidenceH3Survey2020}. 

For example, if AndXIX is produced via high-velocity galaxy collisions, it should be relatively rich in $\alpha$-elements due to vigorous star formation immediately following the collision, during which Type II SN dominate \citep{silkUltradiffuseGalaxiesDark2019}. In contrast, if a significant fraction of AndXIX’s mass has been tidally stripped, it is expected to be more metal-rich than predicted by the present-day stellar mass-metallicity relation for Local Group dwarf galaxies \citep{kirbyUniversalStellarMassStellar2013,kirbyElementalAbundancesM312020}, but should have a comparable $\alpha$-element distribution to other dSph systems of similar original luminosity. 

However, AndXIX’s significant distance \citep[$821^{+32}_{-108}$~kpc:][]{connBayesianApproachLocating2012} makes detailed abundance measurements difficult; the only [Fe/H] abundances determined to date are presented in \citet[henceforth referred to as C20]{C19}. They derive a mean [Fe/H] of $-2.07\pm0.02$ for the galaxy by measuring the strength of the infrared \ion{Ca}{2} triplet (which is empirically correlated with [Fe/H] for RGB stars: e.g.\ \citealt{armandroffMetallicitiesOldStellar1991}) for a coadded spectrum of 81 member stars. However, the low signal-to-noise (S/N) of the individual stars implies a larger uncertainty than the $0.02$~dex implied by their fitting procedure \citepalias{C19}. 

In this paper, we present the first $\alpha$-element abundance measurements for AndXIX, derived from spectral synthesis modelling of coadded medium-resolution spectra. This is a technique that has previously been successfully applied to low-S/N spectra of stars in other M31 satellites \citep{wojnoElementalAbundancesM312020}. We outline the data used and our selection of AndXIX member stars in Section \ref{sec:data}. The details of the abundance measurement procedure, including coaddition of spectra, are given in section~\ref{sec:method}. Section~\ref{sec:results} presents our results, and we discuss their implications for AndXIX’s formation in Section~\ref{sec:origins}. We conclude in Section~\ref{sec:concs}. 

\section{Data} \label{sec:data}
\begin{figure}
	\includegraphics[width=\columnwidth]{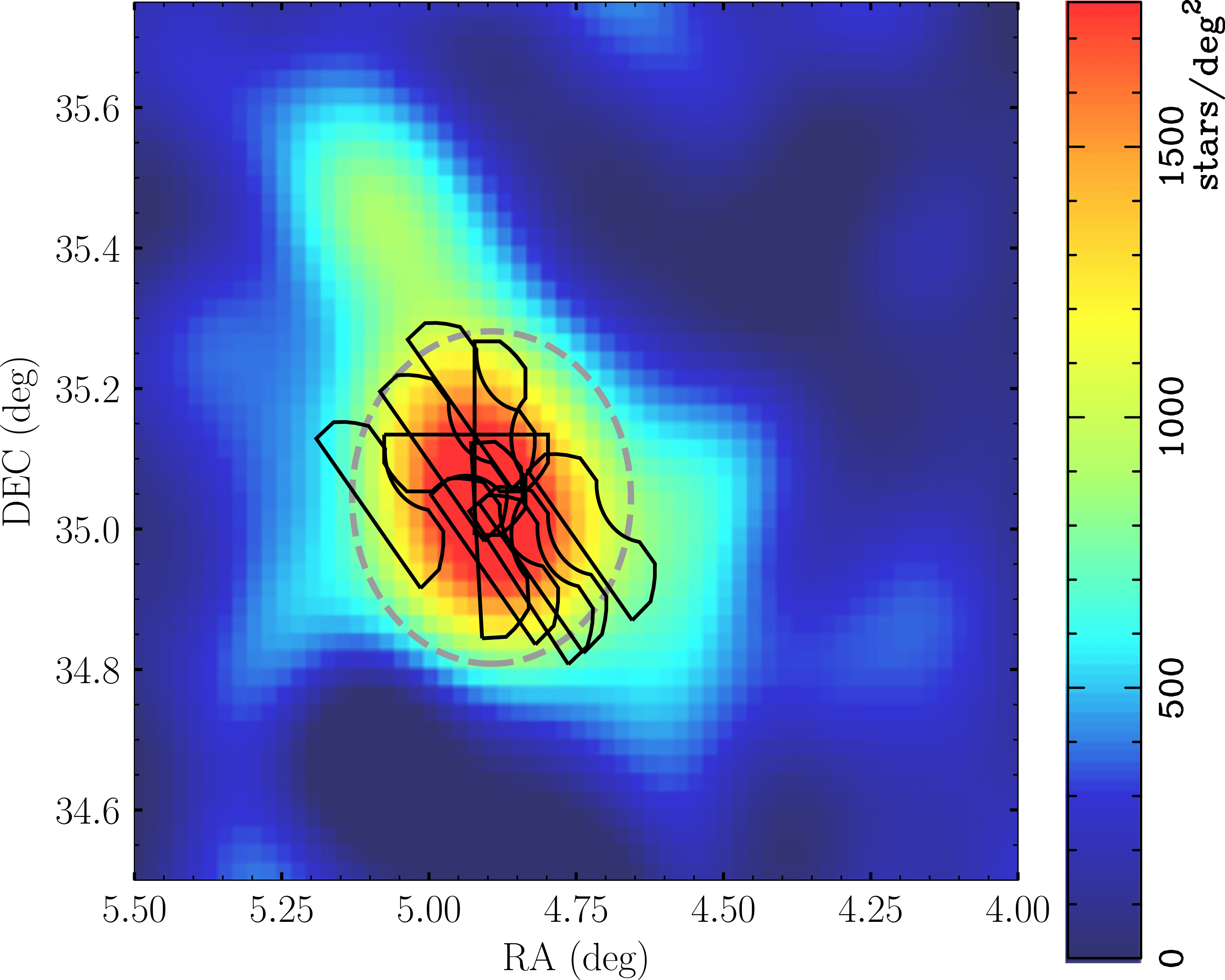}
	\caption{PAndAS density map of AndXIX, taken from figure 2 of \protect\citetalias{C19}. Each black outline approximates the shape of a DEIMOS slitmask used in our analysis (several of which overlap, allowing repeated observations of some stars); the dashed grey ellipse indicates AndXIX's half-light radius \protect\citep[$\sim14$ arcmin:][]{martinPAndASVIEWANDROMEDA2016}.}
	\label{fig:map}
\end{figure}

We utilize observations from a multi-year spectroscopic campaign targeting AndXIX using the Deep Extragalactic Imaging Multi-Object Spectrograph \citep[DEIMOS:][]{faberDEIMOSSpectrographKeck2003} on the Keck-II 10~m telescope, as described in \citetalias{C19}, with the addition of one previously unpublished pilot mask from the same campaign (“d19\_1"). All data are taken with the 1200~l~mm$^{-1}$ grating (R$\sim$6000), with a central wavelength of 8000~\AA, and the OG550 filter, which combined provides wavelength coverage from $\sim6000-9000$~\AA\@. Targets were selected from deep Subaru Suprime-cam imaging as outlined in that paper. We re-reduce the raw spectroscopic data using the \textsc{spec2d} and \textsc{spec1d} pipelines \citep{cooperAstrophysicsSourceCode2012,newmanDEEP2GALAXYREDSHIFT2013} -- the process of which includes includes flat-fielding, sky subtraction, extraction of 1D spectra, and measurement of the line-of-sight velocity for each object via cross-correlation against template spectra \citep{simonKinematicsUltraFaint2007} -- so that the reduced data outputs are consistent with those that were used to calibrate our abundance measurement pipelines (described in \S\ref{sec:method}). Since this reduction is different from that of \citetalias{C19}, we utilize a somewhat different (though largely overlapping) sample of stars from that of \citetalias{C19}; Table~\ref{tab:masks} (analogous to Table 1 of \citetalias{C19}) presents an overview of the data used in our analysis, and Fig.~\ref{fig:map} presents the on-sky positions of the DEIMOS masks (black rectangles) used in our analysis, overlaid on a PAndAS stellar density map of AndXIX \citep{mcconnachieLargescaleStructureHalo2018}. %Notably, masks 8A19A and 8A19b of C20 are not included in our analysis, as we were unsuccessfully able to reduce these masks using our spectroscopic pipeline. 
Since we are using independent data reductions, we derive our own velocity estimates (\S\ref{sec:vels}) and AndXIX membership probabilities (\S\ref{sec:model}), including removal of foreground contaminants (\S\ref{sec:naiew}) for each star; we discuss differences between our final sample and that of \citetalias{C19} in \S\ref{sec:c19mcomp}.

\begin{deluxetable*}{lccrrrrr} \label{tab:masks}
	%% This is the title of the table.
	\tablecaption{Observations of AndXIX used in our analysis, from \citetalias{C19}. A total of 86 targets are considered likely AndXIX members, comprising 65 unique stars for which 18 have at least two measurements on separate masks.}
	%% The \tablehead gives provides the column headers.  It is currently set up so that the column labels are on the top line and the units surrounded by ()s are in the bottom line. 
	\tablehead{\colhead{Mask Name} & \colhead{RA} & \colhead{DEC} & \colhead{Position angle} & \colhead{Exposure time} & \colhead{\# targets} & \colhead{\# targets with} & \colhead{\# And XIX members} \\ 
		\colhead{} & \colhead{(h:m:s)} & \colhead{(d:m:s)} & \colhead{(deg)} & \colhead{(s)} & \colhead{extracted} & \colhead{usable velocities} & \colhead{} } 
	
	%% All data must appear between the \startdata and \enddata commands
	\startdata
	a19m90\tablenotemark{a}   & 00:19:44.69 & +35:05:34.6 & $-90$ & 3600 & 100 & 48 & 12 \\
	a19p0\tablenotemark{b}	  & 00:19:31.02 & +35:07:41.4 & 0   & 3600 & 103 & 46 & 5  \\
	A19S37\tablenotemark{c}	  & 00:19:15.80 & +34:56:28.3 & 37  & 3600 & 82  & 37 & 5  \\
	A19m1                     & 00:19:51.00 & +35:07:00.7 & 40  & 7200 & 102 & 64 & 17 \\
	A19m2                     & 00:19:10.83 & +34:57:23.7 & 40  & 7200 & 91  & 56 & 10 \\
	A19l1                     & 00:20:17.53 & +35:02:58.1 & 40  & 7200 & 96  & 66 & 9 \\
	A19l2                     & 00:18:51.25 & +35:00:15.4 & 40  & 7200 & 85  & 55 & 3  \\
	A19r1                     & 00:19:39.58 & +35:11:19.2 & 40  & 7200 & 97  & 52 & 2  \\
	A19r2                     & 00:19:30.63 & +34:58:02.8 & 40  & 7200 & 90  & 63 & 17 \\
	d19\_1\tablenotemark{d}	  & 00:19:30.84 & +34:59:11.3 & 4   & 900 & 79  & 29 & 7 
	\enddata
	\tablenotetext{a}{Referred to as 7A19a in \citetalias{C19}.} \tablenotetext{b}{Referred to as 7A19b in \citetalias{C19}.} \tablenotetext{c}{Referred to as 8A19c in \citetalias{C19}.} \tablenotetext{d}{Not published in \citetalias{C19}.}
\end{deluxetable*}

\subsection{Velocity measurements}\label{sec:vels}
We apply a number of corrections to the initial velocity measurement produced from the \textsc{spec1d} reduction in order to obtain the final velocity used in our AndXIX membership model. The first of these is conversion to the heliocentric frame. Additionally, in order to account for possible mis-centering of stars within each slit, which manifests as a velocity offset, we measure the systematic variation of the observed wavelength of the atmospheric A-band absorption feature (at $\sim7600$~\AA) as a function of slit position on each individual mask. We fit this “A-band correction” \citep{sohnExploringHaloSubstructure2007} using a polynomial function for stars which have reliable velocity estimates\footnote{i.e. for which visual inspection of the spectra suggests a reasonable velocity has been found using the cross-correlation process.} on each mask, and subsequently apply it to all stars on the mask \citep[as described by][]{Q22}. In total, we obtain reliable velocity measurements for 516 targets. 

Velocity uncertainties are estimated by summing in quadrature the uncertainty estimated by the velocity cross-correlation routine, and a systematic uncertainty of 2.2~km~s$^{-1}$ \citep[derived from data reduced using the same pipeline as we use:][]{simonKinematicsUltraFaint2007}. This is slightly smaller than the systematic uncertainty floor of 3.2~km~s$^{-1}$ derived by \citetalias{C19} through analysis of stars which have repeated observations across different masks. We do, however, check our velocity uncertainties are reasonable through a simpler analysis of duplicate observations within our dataset; out of a total of 430 unique stars in our sample, 70 have at least two reliable velocity measurements. We calculate the LOS velocity difference between paired observations of each star with duplicate observations, scaled by the associated velocity uncertainty (i.e.\ $\frac{v_{i,1}-v_{i,2}}{\sqrt{\sigma_{vi,1}^2+\sigma_{vi,2}^2}}$). The resulting distribution has a standard deviation of $\sim$1.08, indicating our velocity uncertainties are at worst only mildly underestimated. 

\subsection{AndXIX membership model} \label{sec:model}
As we have an independent data reduction process and velocity determination procedure to that of \citetalias{C19}, we independently calculate the probability of each star being associated with AndXIX, largely following the procedure in Section 3.1 of \citetalias{C19}, which we briefly outline below. Our method only differs in the thresholds used to define AndXIX membership; we note where these differ below, and discuss the effects of these different selections further in \S\ref{sec:c19mcomp} and Appendix~\ref{sec:c19compapp}. 

The total probability of a given star $i$ being associated with AndXIX ($P_{\text{tot}}$) is given by Eq.~\ref{eq:ptot}:
\begin{equation}\label{eq:ptot}
	P_{\text{tot},i} = P_{\text{CMD},i}\times P_{\text{vel},i}\times P_{\text{dist},i}
\end{equation}

Here, $P_{\text{CMD}}$ is calculated based on the star’s position on the dereddened (V--i) colour-magnitude diagram relative to a PARSEC isochrone of age 12~Gyr and [Fe/H]=$-1.8$ \citep{bressanPARSECStellarTracks2012}, shifted to a distance modulus of $m-M=24.75$ \cite{connBayesianApproachLocating2012}. This is the same isochrone used in \citetalias{C19}; while \citetalias{Col22} suggest an isochrone of 13.5~Gyr is potentially more appropriate given this corresponds to the 50\% of the star formation in AndXIX, the difference in position between these two isochrones is small and negligibly affects the calculated probabilities. We calculate the minimum cartesian distance from a star to the isochrone locus ($d_{\text{min}}$), and subsequently calculate $P_{\text{CMD}}$ using Eq.~\ref{eq:pcmd}, where $\sigma_{\text{CMD}}=0.1$ to match \citetalias{C19}. 
\begin{equation}\label{eq:pcmd}
	P_{\text{CMD},i}= \exp\left(\frac{-d_{\text{min}}^2}{2\sigma_{\text{CMD}}^2}\right)
\end{equation}
While the use of an 12~Gyr isochrone may preclude potential younger ($\lesssim6$~Gyr) AndXIX populations as being selected as likely member stars, given \citetalias{Col22} find AndXIX has experienced no star formation within the past $\sim8$~Gyr, we consider it unlikely many, if any, genuine AndXIX members are lost as a result of this selection. We do test excluding $P_{\text{CMD}}$ from the final calculation, but find the additional stars considered as potential members are located far from any RGB locus (regardless of assumed age), and therefore either clear interlopers, or at minimum outside the range for which we can determine photometric parameter estimates (a prerequisite for abundance determination in our method: see \S\ref{sec:photometry}). 

$P_{\text{dist}}$ is calculated based on the star’s on-sky location, such that stars closer to the centre of the dwarf are given a higher probability of being associated with it. We first calculate the distance $R$ of each star, in arcmin, from the centre of AndXIX (taken as 0h19m34.5s, 35d02m41.5s from \citetalias{C19}), factoring in the (non-spherical) shape of AndXIX, using Eq.~\ref{eq:radius} \citep[Eq.~5 of][]{martinPAndASVIEWANDROMEDA2016}. 
\begin{eqnarray}\label{eq:radius}
	R_i=\Biggl(\left[\frac{1}{1+\epsilon}\times(x_i\cos(\theta)-y_i\sin(\theta))\right]^2 \\
	+ (x_i\sin(\theta)+y_i\cos(\theta))^2\Biggr)^{0.5} \nonumber
\end{eqnarray}

Here, $x$ and $y$ are given by Eqs.~\ref{eq:xypos}-\ref{eq:xypos2}, where $\alpha_i,\delta_i$ are the on-sky positions of the star, and $\alpha_0,\delta_0$ are the on-sky coordinates of AndXIX's centre.
\begin{eqnarray}
	x_i =& -\cos\left(\delta_i\right)\sin\left(\alpha_i - \alpha_0\right) \label{eq:xypos}\\
	y_i =& \sin\left(\delta_i\right)\cos\left(\delta_0\right) - \cos\left(\delta_i\right)\sin\left(\delta_0\right)\cos\left(\alpha_i - \alpha_0\right) \label{eq:xypos2}
\end{eqnarray}

Here, $\theta$ and $\epsilon$ are structural parameters which describe AndXIX’s shape; $\epsilon$ describes the ellipticity of the system and is linked to the ratio of the major ($a$) and minor ($b$) axes via $\epsilon=1-b/a$, and $\theta$ describes the position angle of the system's major axis, measured east of north. We also adjust AndXIX’s half-light radius for these parameters using Eq.~\ref{eq:rhalf}. We hold these values fixed, with $\theta=34$~deg, $\epsilon=0.58$, and $r_{\text{half}}=14.2$~arcmin \citep{martinPAndASVIEWANDROMEDA2016}.
\begin{equation}\label{eq:rhalf}
	r_h = r_{\text{half}}\times\frac{1-\epsilon}{1+\epsilon\cos\left(\theta\right)}
\end{equation}

$P_{\text{dist}}$ is subsequently calculated using Eq.~\ref{eq:pdist}. However, we note that as the half-light radius of AndXIX is relatively large compared to the on-sky area covered by a single DEIMOS spectroscopic mask (as in Fig.~\ref{fig:map}), $P_{\text{dist}}$ is relatively uninformative and excluding it in the final calculation does not change which stars are considered likely AndXIX members. 
\begin{equation}\label{eq:pdist}
	P_{\text{dist},i} = \exp\left(\frac{-R_i^2}{2r_h^2}\right)
\end{equation}

$P_{\text{vel}}$ is calculated based on the corrected LOS velocity of the stars, accounting for the underlying contamination of both MW foreground stars, and the extended halo of M31. We fit the total LOS velocity distribution of our sample as the sum of three Gaussian components, per Eqs.~\ref{eq:manyprob}-\ref{eq:manyprob3}. Here, $\eta$ describes the fraction of stars in the sample associated with each of the three components, such that $\eta_{\text{A19}}+\eta_{\text{MW}}+\eta_{\text{M31}}=1$. 
\begin{eqnarray}
	P_{\text{A19},i} &= \frac{\eta_{\text{A19}}}{\sqrt{2\pi\left(\sigma_{v,i}^2 + \sigma_{0,\text{A19}}^2\right)}}
	\exp\left(\frac{-0.5(v_i - \mu_{\text{A19}})^2}{\sigma_{v,i}^2 + \sigma_{0,\text{A19}}^2}\right) \label{eq:manyprob}\\ 
	P_{\text{MW},i} &= \frac{\eta_{\text{MW}}}{\sqrt{2\pi\left(\sigma_{v,i}^2 + \sigma_{0,\text{MW}}^2\right)}}
	\exp\left(\frac{-0.5(v_i - \mu_{\text{MW}})^2}{\sigma_{v,i}^2 + \sigma_{0,\text{MW}}^2}\right)  \label{eq:manyprob2}\\ 
	P_{\text{M31},i} &= \frac{\eta_{\text{M31}}}{\sqrt{2\pi\left(\sigma_{v,i}^2 + \sigma_{0,\text{M31}}^2\right)}}
	\exp\left(\frac{-0.5(v_i - \mu_{\text{M31}})^2}{\sigma_{v,i}^2 + \sigma_{0,\text{M31}}^2}\right)  \label{eq:manyprob3}
\end{eqnarray}

While a single Gaussian is an oversimplification of the velocity distribution of MW stars, given the large distance and correspondingly relatively small on-sky area covered by AndXIX, it is sufficient for our purposes of probabilistically identifying AndXIX member stars. The overall likelihood function, is then given by Eq.~\ref{eq:loglikelihood}.
\begin{equation}\label{eq:loglikelihood}
	\log(\mathcal{L})=\sum_i\log\left(P_{\text{A19},i}+P_{\text{MW},i}+P_{\text{M31},i}\right)
\end{equation}

We use \textsc{emcee} \citep{foreman-mackeyEmceeMCMCHammer2013} to sample the posterior distribution of the model parameters and maximize the log-likelihood in Eq.~\ref{eq:loglikelihood}, using uniform priors (given in Table 2 of \citetalias{C19}) for each parameter. We then normalize the probability of a star belonging to any of these three components using Eq.~\ref{eq:normprob} to obtain $P_{\text{vel}}$, taking the 50th percentile of the resulting parameter distributions in the calculation. 
\begin{equation}\label{eq:normprob}
	P_{\text{vel},i} = \frac{P_{\text{A19},i}}{P_{\text{A19},i}+P_{\text{MW},i}+P_{\text{M31},i}}
\end{equation}

\begin{figure*}
	\includegraphics[width=\columnwidth]{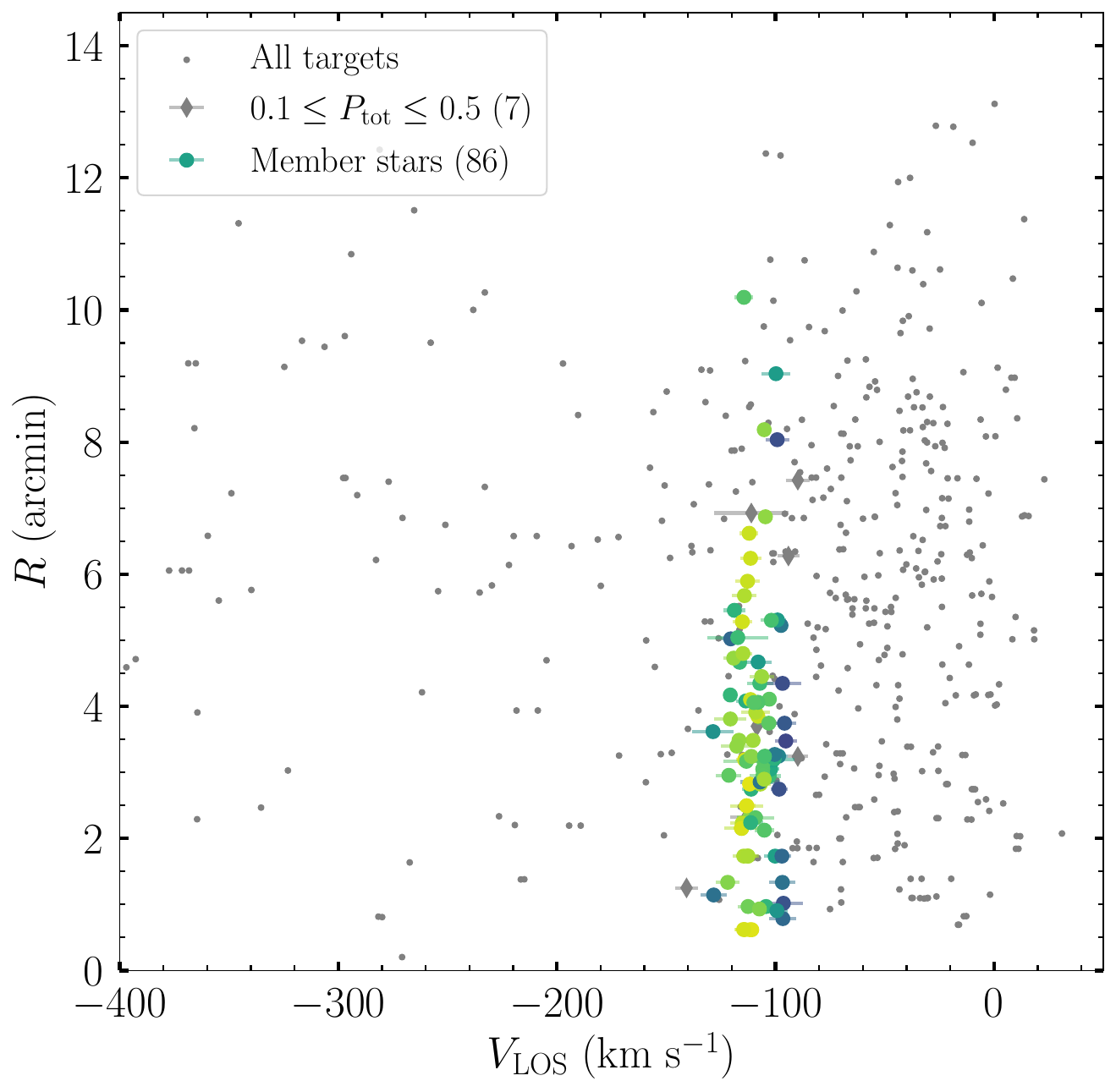}
	\includegraphics[width=1.06\columnwidth]{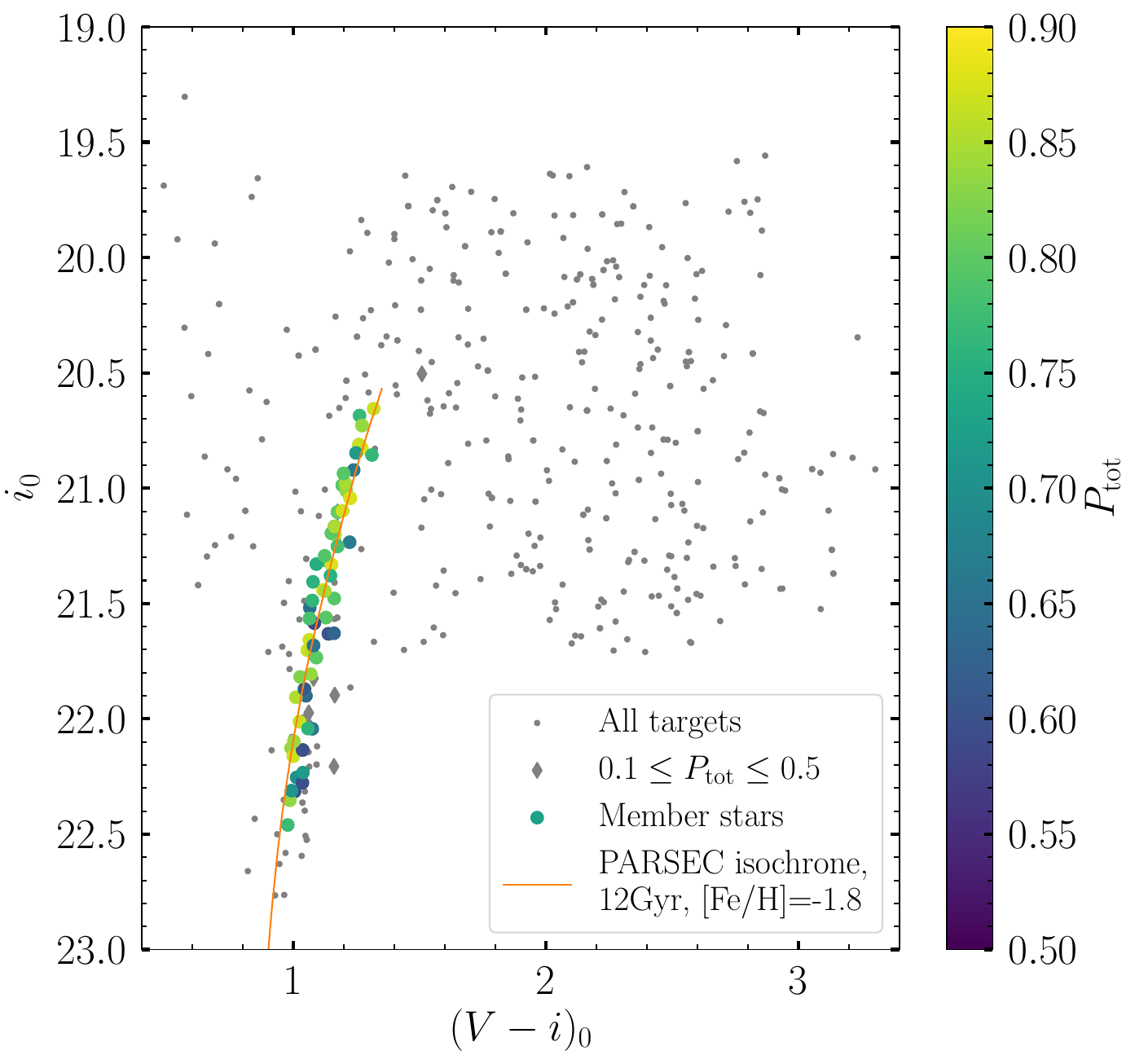}
	\includegraphics[width=\columnwidth]{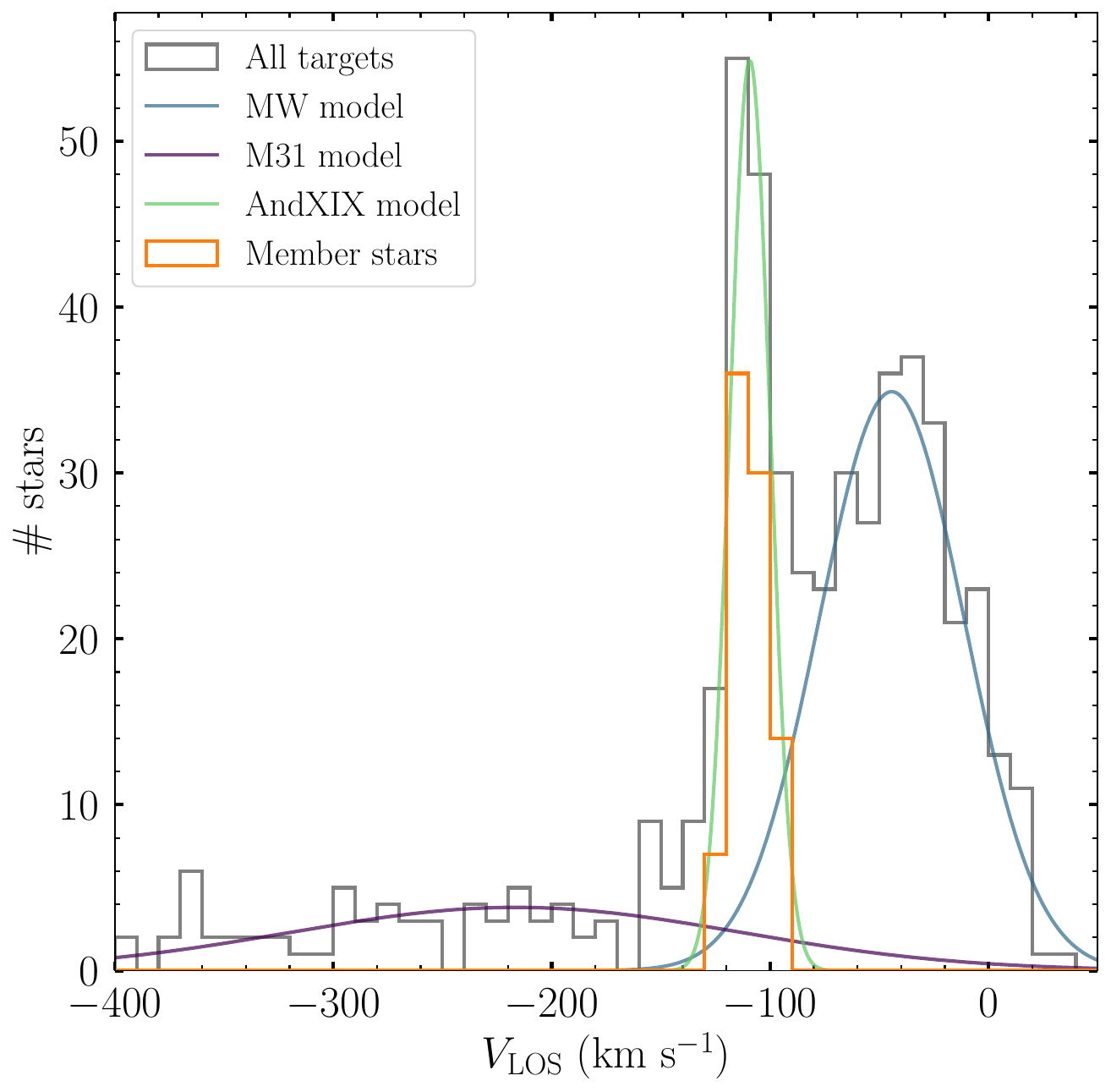}
	\includegraphics[width=\columnwidth]{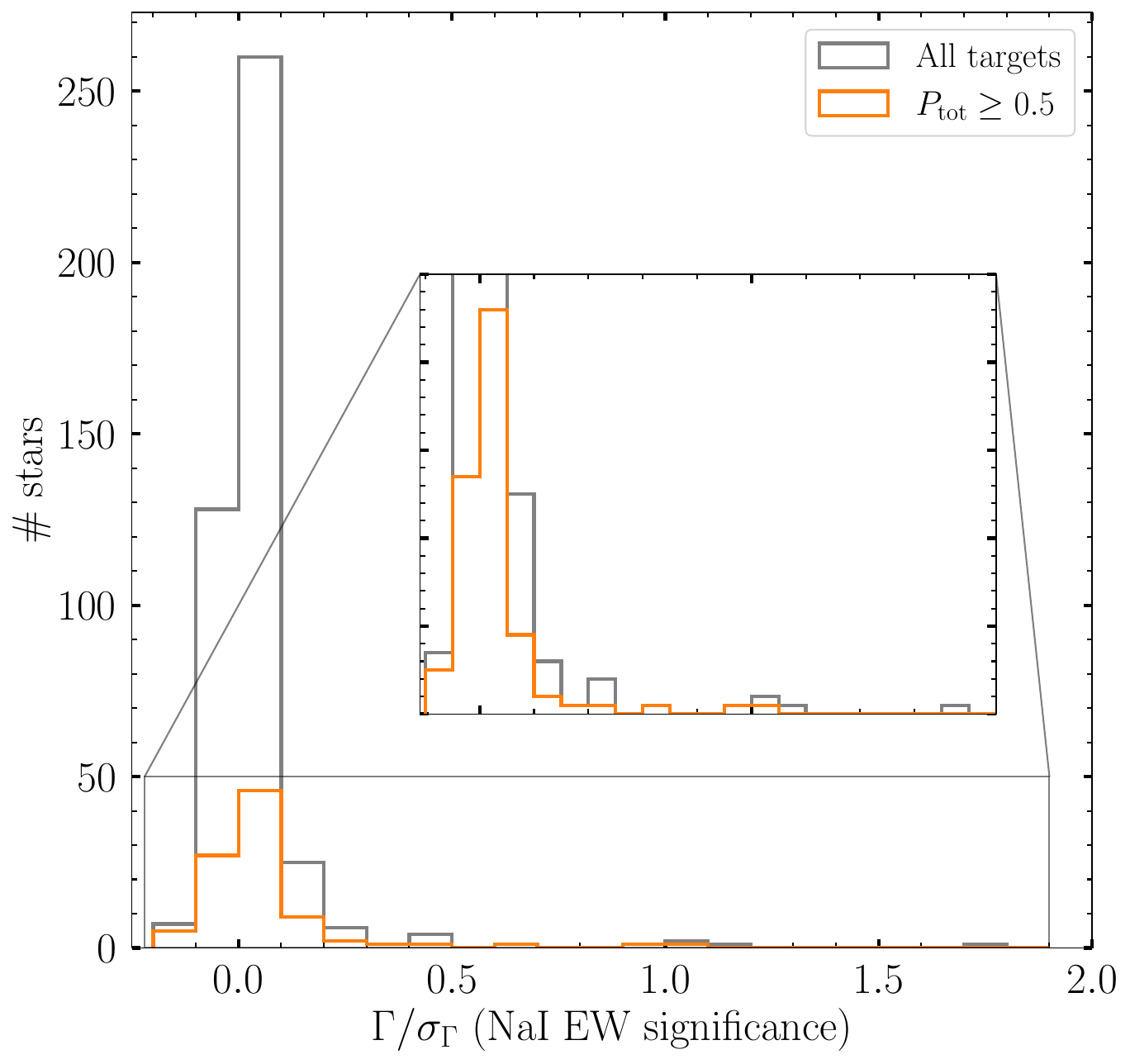}
	\caption{\textit{Top left:} LOS velocity of all targets as a function of distance $R$ in arcmin from the centre of AndXIX. AndXIX member stars have velocity uncertainties indicated with errorbars. \textit{Top right:} Subaru Suprime-cam CMD of all targets. The thin orange line indicates the old, metal-poor PARSEC isochrone used to calculate $P_{\text{CMD}}$. In both upper panels, AndXIX member stars (which pass our $\Gamma/\sigma_\Gamma$ cut: see \S\ref{sec:naiew}) are colour-coded by their total membership probability, based on CMD position (Eq.~\ref{eq:pcmd}), LOS velocity (Eq~\ref{eq:normprob}, and on-sky position (Eq.~\ref{eq:pdist}). Stars with membership probabilities $0.1\leq P_{\text{tot}}\leq0.5$ are indicated by diamond-shaped points. \textit{Bottom left:} LOS velocity histogram for all targets (grey) and AndXIX member stars (orange). Even in the full sample, a clear peak around AndXIX's systemic LOS velocity ($\sim-110$~km~s$^{-1}$) is observed. Thin coloured lines indicate the fitted velocity distributions of AndXIX (green), as well as potential contaminant populations from the MW (blue) and M31 (purple) used to derive $P_{\text{vel}}$. \textit{Bottom right:} Histogram of \ion{Na}{1} doublet detection significance (i.e.\ $\Gamma/\sigma_\Gamma$) for all targets (grey) and targets with $P_{\text{tot}}>0.5$ (orange). We impose a cut at $\Gamma/\sigma_\Gamma<0.2$ to select likely AndXIX member stars (see \S\ref{sec:naiew}).} 
	\label{fig:membermodel}
\end{figure*}

We choose to classify any stars which have $P_{\text{tot}}>0.5$ as potential members of AndXIX. This is a somewhat stricter cut than \citetalias{C19}, who use a threshold of $P_{\text{tot}}>0.1$ to select AndXIX members; we discuss the effect of this difference in \S\ref{sec:c19mcomp}. There are a total of 98 targets we consider to be potential members, encompassing 76 unique stars including 19 for which we have at least two independent measurements on different masks. We demonstrate the results of the membership selection in Fig.~\ref{fig:membermodel} (cf.\ Figures 3 and 4 of \citetalias{C19}). The top right panel shows the Suprime-cam colour-magnitude diagram of the stars, with the thin orange line indicating the PARSEC isochrone used to derive $P_{\text{CMD}}$. The top left panel plots the velocity distribution for all stars as a function of their radial distance from the centre of AndXIX. In both panels, stars that are considered potential AndXIX members are colour-coded by their membership probability $P_{\text{tot}}$, while non-members are indicated as grey dots. The bottom left panel presents a LOS velocity histogram of all targets (grey) and those which we classify as likely members (orange); we overlay the best-fitting models of AndXIX (green), M31 (purple), and the MW (blue) used to derive $P_{\text{vel}}$. 

\subsection{Identifying foreground contaminants}\label{sec:naiew}
In addition to isolating likely AndXIX member stars through the above probability-based cuts, we remove further potential MW contaminants by measuring the equivalent width (EW) of the surface-gravity-sensitive \ion{Na}{1} doublet at 8183 and 8195~\AA: stars with strong \ion{Na}{1} absorption are more likely to be MW foreground dwarfs compared to distant AndXIX giants \citep[e.g.][]{gilbertNewMethodIsolating2006}. We measure the EW of the two \ion{Na}{1} lines in windows of 8180-8190 \AA\ and 8190-8200 \AA, and sum these to obtain a total \ion{Na}{1} EW measurement ($\Gamma$). We bootstrap this measurement by resampling each pixel in these two windows within its uncertainty, and measuring the associated EW of the lines 1000 times; we take the median and standard deviation of the resulting distribution as our final $\Gamma$ value and its uncertainty respectively. In some cases, $\Gamma<0$; this occurs in low S/N spectra where there are no distinct absorption lines in our measurement windows, and instead the mean flux values are above that of the nominal continuum level.

We use the “significance” of the \ion{Na}{1} detection, $\Gamma/\sigma_\Gamma$, as an additional membership criterion. \citetalias{C19} apply a comparable criterion of $\Gamma<2$ to remove potential foreground contaminants and identify AndXIX members; we discuss the effect of this different criterion further in \S\ref{sec:c19mcomp}. The bottom right panel of Fig.~\ref{fig:membermodel} shows the $\Gamma/\sigma_\Gamma$ distribution for the entire sample (grey) and just those stars with $P_{\text{tot}}>0.5$ (orange). The majority of likely member stars have $\Gamma/\sigma_\Gamma\leq0.2$, in a fairly symmetric distribution; we therefore exclude stars that do not pass this criterion from our sample of AndXIX members. While a “significance” of 0.2 is relatively low to consider a detection, this is driven by the very large uncertainties on the $\Gamma$ measurement for all stars due to their low underlying S/N; even spectra where the \ion{Na}{1} doublet is clearly visible are sometimes only detected at a significance of 0.5 or less. Visual inspection confirms that none of the likely AndXIX members which do pass the $\Gamma/\sigma_\Gamma$ cut have strong \ion{Na}{1} absorption. 

In total, 86 targets pass the $\Gamma/\sigma_\Gamma$ cuts and are considered as likely AndXIX members; these comprise 65 unique stars, 18 of which have at least two independent measurements. When considering stars with repeated observations on different masks, in the vast majority of cases all observations are classified consistently. We find only one star for which membership disagrees between observations, due to differing velocity measurements. As both observations have low S/N, and it is not clear that either is more correct than the other, we exclude this star from further analysis as a potential member.  

\subsection{Sample comparison to \citetalias{C19}}\label{sec:c19mcomp}
Our sample of AndXIX members is different to that of \citetalias{C19} due to several factors. In addition to a different underlying target sample, due to the differing reduction methods for the spectroscopic data, we additionally use different selection criteria to define AndXIX membership. We briefly discuss the resulting differences in our samples here, with a more detailed analysis in Appendix \ref{sec:c19compapp}. 

There are 7 targets which have total membership probabilities $0.1\leq P_{\text{tot}}\leq0.5$ and which pass our $\Gamma/\sigma_\Gamma$ cut which we do not consider members, but which would be classified as members according to the $P_{\text{tot}}$ membership criteria in \citetalias{C19}. We define these stars as a “low probability” subsample, and coadd them separately in order to minimise any potential effects from non-member contamination on our primary sample of likely AndXIX members. There are 8 targets that we classify as AndXIX members but are rejected by \citetalias{C19} on the basis of $P_{\text{tot}}<0.1$ due to differing velocity measurements for the stars. These spectra are relatively noisy, and it seems plausible that the different reduction pipelines simply result in different derived velocities for these targets. However, as manual inspection suggests our derived velocities are reasonable, we maintain our classification of these targets as potential AndXIX members. 

We additionally classify 12 targets as potential AndXIX members that \citetalias{C19} exclude on the basis of a \ion{Na}{1} EW measurement exceeding a $\Gamma<2$ cutoff, which is slightly different to the $\Gamma/\sigma_\Gamma>0.2$ we use. Six of these targets are duplicate observations, where the other observation either passes the \citetalias{C19} $\Gamma<2$ cutoff, or is not successfully reduced by their pipeline. We isolate those 18 targets\footnote{the 12 targets and the 6 corresponding duplicate observations.} into a third group -- our “C20 $\Gamma$ difference” sample -- and, like our “low probability” sample, coadd these stars separately to avoid any potential bias of results. \citetalias{C19} also classify an additional three targets as likely members which we do not; we exclude two targets based on $P_{\text{tot}}<0.1$ and one based on $\Gamma/\sigma_\Gamma>0.2$. In all cases these are relatively low S/N spectra, which we speculate produce different results given the different reduction pipelines.

Table~\ref{tab:finalnums} summarises the number of stars in each of our AndXIX member subsamples, and Table~\ref{tab:alldat} presents the properties of all targets for which we can measure a radial velocity, including their membership probabilities, and photometric stellar parameters ($T_{\text{eff,phot}}$, $\log(g)_{\text{phot}}$, and [Fe/H]$_{\text{phot}}$) for potential AndXIX members in any subsample. A small number of targets do not have measured \ion{Na}{1} equivalent widths, where this measurement failed due to a combination of very low S/N and the presence of poorly subtracted sky lines. These targets generally lie outside the range of the isochrone grid we use to derive photometric parameter estimates (see \S\ref{sec:photometry}), and therefore could not be used for subsequent abundance derivation even if they are potentially genuine AndXIX members.

\begin{deluxetable*}{lrrr} \label{tab:finalnums}
	%% This is the title of the table.
	\tablecaption{Summary of likely AndXIX members in this analysis.}
	%% The \tablehead gives provides the column headers.  It is currently set up so that the column labels are on the top line and the units surrounded by ()s are in the bottom line. 
	\tablehead{\colhead{Group} & \colhead{\# Targets} & \colhead{\# Unique stars} & \colhead{\# Stars with duplicate observations}} 
	%% All data must appear between the \startdata and \enddata commands
	\startdata
	$P_{\text{tot}}>0.5$ & 98 & 76 & 19 \\
	Likely AndXIX members (i.e.\ $P_{\text{tot}}>0.5$ \& $\Gamma/\sigma_\Gamma<0.2$) & 86 & 65 & 18 \\
	Primary subsample & 68 & 53 & 12 \\
	C20 $\Gamma$ difference subsample & 18 & 12 & 6 \\
	Low-probability subsample & 7 & 7 & 0 \\ \hline
	\multicolumn{4}{c}{In final coadds} \\ \hline
	Primary subsample & 61 & 49 & 12 \\
	C20 $\Gamma$ difference subsample & 13 & 9 & 4 \\
	Low-probability subsample & 5 & 5 & 0 \\
	\enddata
\end{deluxetable*}

{ \setlength{\tabcolsep}{3pt} 
	\begin{deluxetable*}{lrrlRRRRRRRR} \label{tab:alldat}
		%% This is the title of the table.
		\tabletypesize{\scriptsize}
		\tablecaption{AndXIX target catalogue.}
		%% The \tablehead gives provides the column headers.  It is currently set up so that the column labels are on the top line and the units surrounded by ()s are in the bottom line. 
		\tablehead{\colhead{ID} & \colhead{RA} & \colhead{DEC} & {Mask} & \colhead{V-mag} & \colhead{i-mag} & \colhead{$V_{\text{LOS}}$} & \colhead{$P_{\text{tot}}$} & \colhead{$\Gamma$} & \colhead{$T_{\text{eff,phot}}$} & \colhead{$\log(g)_{\text{phot}}$} & \colhead{[Fe/H]$_{\text{phot}}$} \\ 
			\colhead{} & \colhead{(h:m:s)} & \colhead{(d:m:s)} & \colhead{} & \colhead{} & \colhead{} & \colhead{(km~s$^{-1}$)} & \colhead{} & \colhead{} & \colhead{(K)} & \colhead{(cm~s$^{-2}$)} & \colhead{} } 
		%% All data must appear between the \startdata and \enddata commands
		\startdata
		and19\_00026 & 0:19:22.69 & +35:02:15.3 & a19m90 & $22.331\pm0.01$1 & $21.166\pm0.006$ & $-100.14\pm4.67$ & 0.73 & $-2.3\pm28.1$ & $4511\pm54$ & $0.86\pm0.06$ & $-1.95\pm0.24$ \\
		a19\_00395 & 0:18:51.46 & +35:00:37.2 & A19l2 & $23.575\pm0.032$ & $22.629\pm0.017$ & $-148.20\pm7.52$ & 0.06 & $7.4\pm6.8$ & & & \\
		1233  & 0:19:24.95 & +34:58:44.1 & d19\_1 & $22.461\pm0.013$ & $21.497\pm0.007$ & $-115.22\pm7.35$ & 0.30 & & & & \\
		\enddata
		% \tablenotetext{a}{}
		\tablecomments{V-mag and i-mag are extinction-corrected Vega magnitudes. $P_{\text{tot}}$ is the combined probability of a star being associated with AndXIX (see \S\ref{sec:model}). $\Gamma$ is the equivalent width of the \ion{Na}{1} doublet (see \S\ref{sec:naiew}) where this is possible to measure. $T_{\text{eff,phot}}$, $\log(g)_{\text{phot}}$ and [Fe/H]$_{\text{phot}}$ are derived from a grid of PARSEC isochrones assuming an age of 13~Gyr (see \S\ref{sec:photometry}); these are only included for stars considered potential AndXIX members. This table is available in its entirety in a machine-readable form in the online journal. A portion is shown here for guidance regarding its form and content.} 
\end{deluxetable*} }

\section{Spectroscopic Abundance Measurements}\label{sec:method}
Our overall abundance determination method generally follows that in \citet{wojnoElementalAbundancesM312023}. We first derive photometric estimates (\S\ref{sec:photometry}) of effective temperature ($T_{\text{eff,phot}}$), surface gravity ($\log(g)_{\text{phot}}$), and metallicity ([Fe/H]$_{\text{phot}}$), and use these as initial estimates in the pipeline developed by \citet{escalaElementalAbundancesM312019} to obtain [Fe/H] and [$\alpha$/Fe] measurements for each individual stellar spectrum by comparison to a grid of synthetic spectra (\S\ref{sec:synthspec}). However, as our targets are faint, we subsequently coadd spectra as in \citet{wojnoElementalAbundancesM312020, wojnoElementalAbundancesM312023} and measure average [Fe/H] and [$\alpha$/Fe] abundances for groups of several stars (\S\ref{sec:coadds}). Previous work \citep{wojnoElementalAbundancesM312020} has shown these coadded measurements accurately reflect the weighted average of the underlying values of the individual contributing component stars, and that the method provides comparable abundance distributions to those derived from individual stars using both low- \citep{wojnoElementalAbundancesM312020} and high-resolution spectroscopy \citep{escalaElementalAbundancesM312019}.

\subsection{Photometric parameter estimates}\label{sec:photometry}
We compare the dereddened Suprime-cam photometry of each individual star to a grid of PARSEC isochrones \citep{bressanPARSECStellarTracks2012}\footnote{version 3.6, accessed via \url{http://stev.oapd.inaf.it/cgi-bin/cmd}} in order to derive photometric stellar parameter estimates. We assume an age of 13~Gyr as this is close to the 13.5~Gyr at which 50\% of the star formation in AndXIX is complete \citepalias{Col22}; at ages older than 13~Gyr, a handful of stars are located beyond the low-metallicity bound of the isochrones ([M/H]=$-2.2$) and thus do not have reliable stellar parameter estimates. During subsequent spectral abundance measurements, the effective temperature and surface gravity for each star are held fixed at the values derived from photometry. While using a single isochrone age will impact the derived photometric parameters due to the age-metallicity degeneracy, we find varying the isochrone age between 10-14~Gyr has a minimal effect on our results (see \S\ref{sec:resmain}); and, per \S\ref{sec:model}, it is unlikely that there are significantly younger populations in AndXIX for which the derived photometric parameters are significantly inaccurate.

We note that the PARSEC isochrones used are solar-scaled (i.e. assume [$\alpha$/Fe]=0), which can potentially affect the derived photometric parameters and therefore the resulting spectroscopic abundances. However, \citet{kirbyMetallicityAlphaElement2008} and \citet{vargasDistributionAlphaElements2014} -- both of whom use very similar methods to that of our analysis -- find that photometric effective temperature and surface gravity are largely insensitive to using $\alpha$-enhanced isochrones up to [$\alpha$/Fe]=0.3-0.4, and this therefore negligibly affects the derived spectroscopic abundances.

Despite the relatively narrow CMD concentration of likely AndXIX members, as seen in the upper right panel of Fig.~\ref{fig:membermodel}, the photometric metallicities we derive for these stars show a moderate spread, spanning a range $-2.2\lesssim$[Fe/H]$\lesssim-1.4$. This is comparable to the range of metallicities later derived using our spectroscopic pipeline (\S\ref{sec:results}). Stars in the "low-probability" subsample span a comparably wide metallicity range, with 1-2 stars reaching photometric metallicities as metal-rich as [Fe/H]$\sim-1$. This may indicate these stars are potential interlopers from the MW or M31. However, as these values are only used as an initial estimate for our spectroscopic pipeline, which provides better constraints, we proceed with their analysis. 

\subsection{Synthetic spectra}\label{sec:synthspec}
To obtain abundance estimates for the stars, we compare the observed 1D spectra to a grid of synthetic spectra generated by the spectral synthesis code \textsc{MOOG} \citep{snedenStarToStarAbundanceVariations1997}, utilizing ATLAS9 stellar atmospheric models \citep[and references therein]{kirbyGridsATLAS9Model2011}. The mixing length parameter for these models is $l/H_p=1.25$. The associated line list draws from the Vienna Atomic Line Database \citep{kupkaVALDProgressVienna1999}, as well as molecular lines \citep{kuruczAtomicMolecularData1992} and hyperfine transitions \citep{kuruczAtomicDataInterpreting1993} which are tuned to match the line strengths of the Sun and Arcturus \citep{kirbyGridsATLAS9Model2011,escalaElementalAbundancesM312019}. 

Synthetic spectra are generated across linearly spaced parameter ranges of $3500\leq T_{\text{eff}}$~(K)$\leq8000$, $0.0\leq\log(g)\leq5.0$, $-5.0\leq$[Fe/H]$\leq0.0$, and $-0.8\leq[\alpha$/Fe]$\leq1.2$. The microturbulent velocity $\xi$ used is a linear interpolation between two values out of 0, 1, 2, and 4 km~s$^{-1}$ which bracket that given by the relation $\xi=2.13–0.23\log(g)$ for red giant stars from \citet{kirbyMultielementAbundanceMeasurements2009}. They cover a wavelength range of $6300-9100$~\AA, and have a wavelength spacing of 0.02~\AA\@. The synthetic spectra are then interpolated onto the observed wavelength array and smoothed to match the observed spectral resolution ($\Delta\lambda$). While the spectral resolution is known to slowly vary both as a function of wavelength\footnote{Per \citet{escalaElementalAbundancesM312019}, assuming a fixed value of $\Delta\lambda$ with wavelength has no net effect on the derived abundances, and at most increases their associated statistical uncertainties.} and also between masks \citep{escalaElementalAbundancesM312020}, we keep this fixed at $\Delta\lambda=0.45$ to permit the coaddition of stars across different masks \citep{wojnoElementalAbundancesM312023}. 

\subsection{Coadding spectra}\label{sec:coadds}
As our AndXIX spectra have relatively low S/N, it is not possible to derive individual abundance estimates for each target. Instead, we coadd spectra from several member stars in order to calculate average abundance estimates. We initially coadd multiple exposures of the same star on different masks where these exist; we find for 5 stars this is sufficient to successfully\footnote{Defined later in this section.} derive an abundance estimate for the star. For all other stars (including repeat observations of stars on different masks which do not converge when coadded alone), we sort by photometric metallicity and identify groups of $\sim$5 stars to coadd\footnote{\citet{wojnoElementalAbundancesM312020} find negligible differences in the abundances derived for coadd groups sorted by $T_{\text{eff,phot}}$ or [Fe/H]$_{\text{phot}}$.}. The pipeline we use to measure coadded abundances is very similar to that which would otherwise be used for individual stars as described in \citet{kirbyMetallicityAlphaElement2008} and \citet{escalaElementalAbundancesM312019}, with modifications for the coadded spectra as described in \citet{wojnoElementalAbundancesM312020, wojnoElementalAbundancesM312023}. We briefly summarise the procedure here.

Each individual 1D spectrum in the coadd is shifted to the rest frame using its LOS velocity\footnote{As this is a purely observational shift, we do not apply the heliocentric or A-band velocity corrections in this step.}, and corrected for telluric absorption features as in \citet{kirbyMetallicityAlphaElement2008} by dividing the observed spectrum by the scaled template spectrum of a spectrophotometric standard star \citep{simonKinematicsUltraFaint2007}. An initial continuum normalisation is performed by fitting a third-order B-spline function with a breakpoint spacing of 100 pixels to continuum regions of the spectrum that are not strongly affected by absorption lines, as defined by \citet{kirbyMetallicityAlphaElement2008}. The normalised spectra for all stars in the coadd group are then combined per Eq.~\ref{eq:coaddflux}, with the flux in each pixel ($f_i$) weighted by the inverse variance per pixel ($1/\sigma_i$) for each of the $n$ spectra in the group. The associated uncertainty for each pixel in the resultant coadded spectrum ($\sigma_i$) is therefore given by Eq.~\ref{eq:coaddunc}. 

\begin{equation}\label{eq:coaddflux}
	f_i = \frac{\sum_j^nf_{i,j}/\sigma_{i,j}^2}{\sum_j^n1/\sigma_{i,j}^2}
\end{equation}

\begin{equation}\label{eq:coaddunc}
	\sigma_i = \left(\sum_j^n\frac{1}{\sigma_{i,j}^2}\right)^{-1/2}
\end{equation}

We subsequently perform an initial fit to the coadded spectrum to determine a first estimate of [Fe/H], during which [$\alpha$/Fe] is held fixed at zero. The synthetic spectrum with which the coadd is compared during this process is itself a coaddition of several synthetic spectra: each observed spectrum in the group has a corresponding synthetic spectrum, which has $T_{\text{eff}}$ and $\log(g)$ fixed at the corresponding photometric values for the associated observed spectrum. When coadding the synthetic spectra, we weight each individual synthetic spectrum by the same inverse variance array as the corresponding observed spectrum in order to account for the differing weight each observed spectrum contributes to the final coadd. In this and all subsequent fitting steps, we use a Levenberg-Marquardt algorithm to minimise the difference between the coadded synthetic and observed spectra, weighting the comparison by the inverse variance of the coadded observed spectrum. 

We then perform a fit to the coadded spectrum to derive an initial estimate of [$\alpha$/Fe] in the same manner, during which [Fe/H] is held fixed at the previously determined value. We fit only spectral regions shown to be sensitive to the $\alpha$-elements of Mg, Si, Ca, and Ti as in \citet{escalaElementalAbundancesM312019} and \citet{kirbyMetallicityAlphaElement2008}; additionally, we mask the infrared \ion{Ca}{2} triplet as our synthetic spectra cannot accurately reproduce the shape of these lines \citep{kirbyMetallicityAlphaElement2008}. Strong TiO features are also not well-reproduced by our synthetic spectral grid, but visual inspection of the spectra confirms none of our AndXIX member stars have these features. High-resolution studies of other dSphs reveal that abundances of different individual $\alpha$-elements can vary within stars \citep[see, e.g.,][]{hillVLTFLAMESHighresolution2019,thelerChemicalEvolutionDwarf2020}, though these generally follow similar overall trends \citep{kirbyElementalAbundancesM312020}. However, our measurements have sufficiently low S/N that even in the coadded groups, it is not possible to measure separate abundances for the individual $\alpha$ elements; the [$\alpha$/Fe] we measure reflects an average of the different contributing elements, and cannot capture these variations.

The continuum of the coadded spectrum is subsequently re-normalised using the best-fitting coadded synthetic spectrum. This sequential fitting process and continuum normalisation is repeated until the derived [Fe/H] and [$\alpha$/Fe] values converge to within tolerances of 0.001~dex. Once continuum refinement is complete, we use the resulting coadded observed spectrum for all subsequent fits. 

We then perform an updated fit for [Fe/H], holding [$\alpha$/Fe] fixed at the last best-fit value. We then fit for the ultimate [$\alpha$/Fe] abundance while [Fe/H] is held constant at that value, and subsequently perform a final fit to determine the ultimate [Fe/H] value while holding [$\alpha$/Fe] constant at its ultimate value. 

Uncertainties in the fitted abundances are also calculated, along with a reduced $\chi^2$ statistic for both [Fe/H] and [$\alpha$/Fe]. These uncertainties are combined in quadrature with systematic uncertainty floors of 0.101~dex for [Fe/H], and 0.084~dex for [$\alpha$/Fe] \citep[derived from M31 data of similar quality and analysed using the same reduction and abundance pipelines:][]{gilbertElementalAbundancesM312019}. We define a coadd as successful (i.e., as having a secure abundance determination) if it meets all of the following criteria: a) the $\chi^2$ contours for both parameters vary smoothly; b) the fit parameters are not located at an edge of the spectral synthesis grid, and c) uncertainties on both [Fe/H] and [$\alpha$/Fe] are less than $0.45$ (including statistical uncertainties). 

We initially identify coadd groups of five targets, starting from those with the lowest photometric metallicity estimates. If a coadd group does not have a secure abundance measurement, we test removing poor-quality or otherwise anomalous spectra from the group and refitting. Should this still not converge, we then add the next closest photometric metallicity target to the coadd group and refit, adjusting subsequent coadd groups accordingly. In total, we obtain 20 unique abundance measurements: 15 in our primary sample, 4 for which we disagree with \citetalias{C19} on membership based on \ion{Na}{1} EWs (our ``C20 $\Gamma$ difference'' subsample), and one for our ``low-probability'' subsample. Table~\ref{tab:finalnums} summarises the number of targets which contribute to the coadds in each of the different subgroups, and Table~\ref{tab:finalprops} presents the properties of the final coadd groups. Coadds with single values for the photometric property range indicate these are comprised solely of duplicate observations of the same star.

\begin{deluxetable*}{lrrr} \label{tab:finalprops}
	%% This is the title of the table.
	\tablecaption{Summary of AndXIX coadd groups.}
	\tabletypesize{\small}
	%% The \tablehead gives provides the column headers.  It is currently set up so that the column labels are on the top line and the units surrounded by ()s are in the bottom line. 
	\tablehead{\colhead{Included Target IDs} & \colhead{[Fe/H]$_{\text{phot}}$ range} & \colhead{$T_{\text{eff,phot}}$ range (K)} & \colhead{$\log(g)_{\text{phot}}$ range (cm~s$^{-2}$)}} 
	%% All data must appear between the \startdata and \enddata commands
	\startdata
	\multicolumn{4}{c}{Primary subsample} \\ \hline
	a19\_00185, 1530, a19\_00061, and19\_00104 & -2.17 -- -2.11 & 4443 -- 4608 & 0.63 -- 0.96 \\
	a19\_00134, a19\_00095, and19\_00022, a19\_00069 & -2.03 -- -1.99 & 4506 -- 4559 & 0.83 -- 0.96 \\
	a19\_00115, a19\_00098, a19\_00098, and19\_00047, 1499 & -2.06 -- -1.96 & 4467 -- 4623 & 0.77 -- 1.07 \\
	and19\_00026, and19\_00026, and19\_00026 & -1.95 & 4511 & 0.86 \\
	and19\_00058, a19\_00172 & -1.95 & 4399 & 0.63 \\
	1239, and19\_00032, a19\_00072, 1505, a19\_00060 & -1.95 -- -1.92 & 4403 -- 4511 & 0.67 -- 0.87 \\
	and19\_00303, a19\_00168, a19\_00067, a19\_00108, a19\_00502 & -1.88 -- -1.85 & 4411 -- 4704 & 0.72 -- 1.32 \\
	a19\_00093, and19\_00020 & -1.83 & 4373 & 0.66 \\
	and19\_00040, a19\_00352, a19\_00077, a19\_00157, and19\_02850 & -1.84 -- -1.81 & 4328 -- 4683 & 0.57 -- 1.3 \\
	a19\_00360, a19\_00183, a19\_00392, a19\_00142, and19\_00283 & -1.81 -- -1.79 & 4576 -- 4621 & 1.10 -- 1.19 \\
	1214, and19\_00028, a19\_00111, a19\_00111 & -1.79 & 4409 & 0.77 \\
	a19\_00043, 1540, a19\_00078, a19\_00443 & -1.79 -- -1.77 & 4463 -- 4683 & 0.88 -- 1.33 \\
	a19\_00343, and19\_00265, a19\_00428 & -1.77 -- -1.73 & 4580 -- 4646 & 1.15 -- 1.26 \\
	a19\_00411, a19\_00431, a19\_00149, and19\_00043, a19\_00087 & -1.68 -- -1.61 & 4361 -- 4697 & 0.82 -- 1.41 \\
	and19\_00134, a19\_00418, a19\_00348, a19\_00546, a19\_00182 & -1.61 -- -1.46 & 4399 -- 4593 & 0.95 -- 1.35 \\ \hline
	\multicolumn{4}{c}{Low-probability subsample} \\ \hline
	a19\_00377, a19\_00302, a19\_00425, a19\_00414, a19\_00384 & -1.67 -- -1.03 & 4322 -- 4576 & 1.08 -- 1.24 \\ \hline
	\multicolumn{4}{c}{C20 $\Gamma$ difference subsample} \\ \hline
	and19\_00057, a19\_00170 & -2.06 & 4499 & 0.78 \\
	and19\_00048, a19\_00156 & -1.85 & 4539 & 0.99 \\
	a19\_00159, a19\_00104, a19\_00104, a19\_00075, a19\_00083 & -2.14 -- -1.67 & 4285 -- 4617 & 0.62 -- 1.12 \\
	a19\_00475, a19\_00403, and19\_00053, a19\_00165 & -1.59 -- -1.56 & 4453 -- 4606 & 1.02 -- 1.29
	\enddata 
	\tablecomments{A repeated target ID indicates the same star is observed twice on different masks. In addition, some duplicate observations of the same star on different masks are associated with different target IDs depending on when the sets of observations were taken; these are indicated by single values in the subsequent columns.} 
\end{deluxetable*}

\section{Results}\label{sec:results}
\subsection{The abundance distribution of AndXIX}\label{sec:resmain}
\begin{figure}
	\includegraphics[width=\columnwidth]{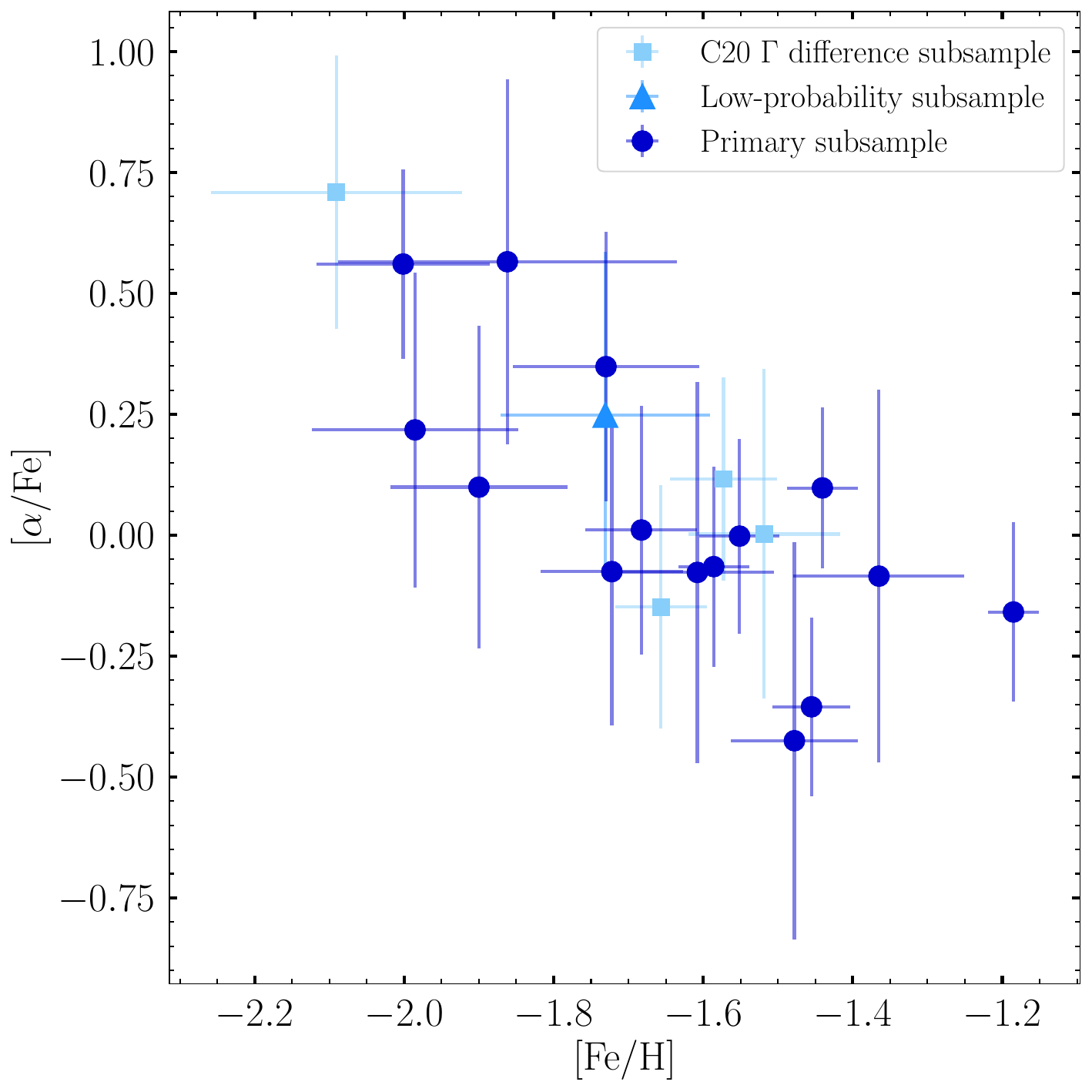}
	\caption{[$\alpha$/Fe] vs.\ [Fe/H] from coadded spectra for AndXIX. Circles represent our primary sample of AndXIX members. Triangular points represent our ``low-probability'' subsample, which comprises stars with $0.1<P_{\text{tot}}<0.5$ and also pass our $\Gamma/\sigma_\Gamma$ cut.  Square points represent our ``C20 $\Gamma$ difference'' subsample, which comprises stars we classify as likely AndXIX members but \citetalias{C19} do not based on differing \ion{Na}{1} EW thresholds.} \label{fig:abund}
\end{figure}

Fig.~\ref{fig:abund} presents the [$\alpha$/Fe]-[Fe/H] abundance distribution for all our successful AndXIX coadds (tabulated in Table~\ref{tab:abunds}); there is a striking decline in [$\alpha$/Fe] as a function of [Fe/H] across the entire metallicity range. The distribution of the “C20 $\Gamma$ difference” (\S\ref{sec:c19mcomp}) and “low-probability” subsamples largely overlap that of the primary subample of high-confidence AndXIX members, and Kolmogorov-Smirnov (K--S) tests also indicate no significant differences in the abundance distributions of these subsamples compared to the primary sample. This indicates these subsamples are likely also comprised predominantly, if not entirely, of genuine AndXIX members, and are not strongly contaminated by non-members from either the M31 or MW halos, both of which have comparatively flatter [$\alpha$/Fe]--[Fe/H] distributions at higher [$\alpha$/Fe] \citep[e.g.][see also \S\ref{sec:ogalcomp}]{escalaElementalAbundancesM312020,mckinnonHALO7DIIIChemical2023}. Accordingly, for the remainder of the analysis, we analyse these collectively (unless otherwise specified).

\begin{deluxetable}{RR} \label{tab:abunds}
	%	\tablewidth{\textwidth}
	%% This is the title of the table.
	\tablecaption{Elemental abundances of coadded groups.}
	%% The \tablehead gives provides the column headers.  It is currently set up so that the column labels are on the top line and the units surrounded by ()s are in the bottom line. 
	\tablehead{\colhead{[Fe/H]} & \colhead{[$\alpha$/Fe]}} 
	%% All data must appear between the \startdata and \enddata commands
	\startdata
	\multicolumn{2}{l}{Primary subsample} \\ \hline
	-2.00\pm0.12 & 0.56\pm0.2 \\	
	-1.99\pm0.14 & 0.22\pm0.33 \\	
	-1.90\pm0.12 & 0.10\pm0.33 \\	
	-1.86\pm0.23 & 0.57\pm0.38 \\	
	-1.73\pm0.12 & 0.35\pm0.28 \\	
	-1.72\pm0.10 & -0.08\pm0.32 \\	
	-1.68\pm0.07 & 0.01\pm0.26 \\	
	-1.61\pm0.10 & -0.08\pm0.39 \\	
	-1.59\pm0.05 & -0.07\pm0.21 \\	
	-1.55\pm0.05 & 0.00\pm0.20 \\	
	-1.48\pm0.08 & -0.43\pm0.41 \\	
	-1.46\pm0.05 & -0.35\pm0.18 \\	
	-1.44\pm0.05 & 0.10\pm0.17 \\	
	-1.36\pm0.11 & -0.08\pm0.39 \\	
	-1.19\pm0.03 & -0.16\pm0.19 \\	\hline	 
	\multicolumn{2}{l}{Low-priority subsample} \\ \hline
	-1.73\pm0.14 & 0.25\pm0.34 \\ \hline
	\multicolumn{2}{l}{C20 $\Gamma$ difference subsample} \\ \hline
	-2.09\pm0.17 & 0.71\pm0.28 \\
	-1.66\pm0.06 & -0.15\pm0.25 \\
	-1.57\pm0.07 & 0.12\pm0.21 \\
	-1.52\pm0.10 & 0.00\pm0.34 \\
	\enddata
\end{deluxetable}

We do test varying the isochrone age used to derive the photometric effective temperature and surface gravity for the individual stars that comprise the coadds in order to assess the resulting effects on the derived coadd abundances. However, for isochrones ranging between 10-14~Gyr (a range that accounts for $\sim90$\% of the star formation in AndXIX: \citetalias{Col22}), the resulting derived abundances change negligibly (by a maximum of 0.04~dex in [Fe/H] and 0.03~dex in [$\alpha$/Fe]: well within the uncertainties on our canonical values). The metallicity range spanned by our coadded measurements is also very similar to that of the photometric metallicities for the underlying sample of individual targets. 

The lack of an [$\alpha$/Fe] plateau at low [Fe/H] indicates that most stars in AndXIX formed after enough SNIa had occurred to depress [$\alpha$/Fe]. While the oldest (and hence most metal-poor) stars should have been formed prior to the SNIa, our measurements do not extend to sufficiently metal poor values to identify these stars and hence the location of the [$\alpha$/Fe] “knee” in AndXIX. 

Several factors likely contribute to this lack of detection. Low-metallicity stars are inherently relatively rare; in studies using high-resolution spectroscopy \citep[e.g.][]{thelerChemicalEvolutionDwarf2020}, the number of stars in low-metallicity [$\alpha$/Fe] plateaus is significantly fewer than the number of stars “post-knee”. In addition, as low-metallicity stars have shallower spectral lines, they are more difficult to derive confident velocities for in relatively low-S/N observations -- including those used in this analysis. As seen in Table~\ref{tab:masks}, we only successfully measure velocities for $\sim50$\% of the targets a given DEIMOS mask; it is plausible that metal-poor stars represent a higher than average fraction of those targets which are “lost” and therefore are not included in our analysis. This may partially explain why [$\alpha$/Fe] “knees” are not observed in similar-luminosity dSphs at metallicity values above [Fe/H]=$-2.5$ in DEIMOS data \citep{kirbyMultielementAbundanceMeasurements2011}, but have been detected at lower metallicities in higher-resolution studies \citep[e.g.][]{lemasleVLTFLAMESSpectroscopy2014,thelerChemicalEvolutionDwarf2020}, which are better able to resolve these shallow lines (at sufficiently high S/N). Finally, the averaging effect introduced by our coaddition of spectra masks the (already inherently narrow) tails of the metallicity distribution, including at the metal-poor end. Deeper observations of AndXIX, which allow abundances to be derived for individual stars, would likely help to identify its [$\alpha$/Fe] knee -- and provide better estimates of the systematic uncertainty associated with our abundance measurements, through duplicate target observations.  %and help confirm that the lack of a low-metallicity plateau in [$\alpha$/Fe] is not due solely due to the effects of coaddition. 

We also do not see an [$\alpha$/Fe] plateau at high metallicity. Such plateaus can occur when star formation is constant for a sufficiently long duration that the ratio of Type Ia to Type II supernovae becomes constant, with a resulting equilibrium in the production of $\alpha$-elements and iron \citep{kirbyMultielementAbundanceMeasurements2011}. Alternatively, they can also form in systems that have very rapid repeated bursts of star formation, which mimics the effects of constant star formation \citep{revazDynamicalChemicalEvolution2009}. However, these plateaus are typically only observed in more massive systems with extended star formation such that they reach metallicities above [Fe/H]$\gtrsim-1.2$ \citep{kirbyMultielementAbundanceMeasurements2011}. Given similar-luminosity systems like Sextans do not show this plateau, and the metallicity distribution of AndXIX does not extend to very metal-rich values, the lack of a high-metallicity plateau is expected. This also aligns with the star formation history (SFH) of AndXIX presented by \citetalias{Col22}, which indicates that two distinct bursts of SF formed $>90$\% of the stars in AndXIX, without the extended star formation or rapid repeated bursts necessary for [$\alpha$/Fe] equilibrium to be reached. 

\subsection{Comparison with literature metallicities}\label{sec:c19rcomp}
While \citetalias{C19} do not derive $\alpha$-element abundances for AndXIX, they do present [Fe/H] estimates derived from calibrations of the equivalent width (EW) of the near-infrared Calcium II triplet (CaT). As we mask the CaT during our usual fitting process due to strong non-LTE effects which our synthetic spectra cannot accurately describe, this effectively provides an entirely independent [Fe/H] estimate. 

Their metallicity distribution, derived for individual AndXIX member stars with S/N$>5$~\AA$^{-1}$, has a mean [Fe/H]$=-1.8\pm0.1$~dex and a dispersion of $\sim0.5$~dex (after accounting for uncertainties in the individual measurements, which themselves are on the order of 0.5~dex). In contrast, the metallicity distribution we derive from our full sample of coadded measurements has a mean [Fe/H] of $-1.50\pm0.02$~dex, and a dispersion, accounting for the individual measurement uncertainties, of $\sim0.2$~dex. Given that we have significantly fewer data points, each of which combines data from multiple stars, it is no surprise the width of our distribution is substantially narrower.

Less clear is the source of the $\sim3\sigma$ higher mean metallicity we derive, as there are several factors which could potentially contribute to this difference. Most likely is that this is due to the two different techniques used to derive the metallicities. In order to test this effect, we also calculate [Fe/H] values for stars we consider potential AndXIX members using the same CaT equivalent calibration as in \citetalias{C19}. To do so, we first fit a Voigt profile to the 8542 and 8662~\AA\ \ion{Ca}{2} lines in the each spectrum using the package \textsc{phew} \citep{nunezPHEWPytHonEquivalent2022}, and sum the associated EW for each. We then use the calibrations presented in \citet{starkenburgNIRCaIi2010} based on the absolute magnitude $M_V$ (which we convert from the observed V-band magnitude assuming a distance modulus of 24.57 per \S\ref{sec:model}) to derive a [Fe/H] value for each star. 

Because the S/N of the spectra are low, we find that we can only reliably fit the CaT lines, and therefore derive a useful metallicity, for 18 stars. The mean of this distribution is $-1.81\pm0.06$; entirely consistent with that derived by \citetalias{C19}. This suggests that the different underlying sample of member stars and data reduction methods we use compared to \citetalias{C19} do not significantly affect the derived metallicities, but that the different techniques used to derive the metallicity do have an impact.

In particular, we point out that the \citet{starkenburgNIRCaIi2010} calibration is derived using model spectra with [$\alpha$/Fe]$\geq0$. Per Fig.~\ref{fig:abund}, this is only true for 7 of our 20 coadds: predominantly those which are metal-poor, with [Fe/H]$\lesssim-1.7$. In contrast, the majority of our measurements -- particularly those which are more metal-rich -- have [$\alpha$/Fe]$<0$. This means that a [Fe/H] estimate derived from the CaT calibration will be underestimated for these (metal-rich) stars. Indeed, for a similar CaT calibration, \citet{dacostaCaIiTriplet2016} finds that a reduction in [$\alpha$/Fe] by 0.5~dex will result in [Fe/H] values derived from CaT abundances being underestimated by $\sim0.15$~dex. Since our data span an even larger [$\alpha$/Fe] range than this, we expect the CaT metallicities to be similarly affected, and we suggest this is largely why the average metallicity we derive for AndXIX using spectral synthesis is higher than that derived using the CaT.

Another potential contributor to the difference in mean metallicities is that our abundances are derived for coadded spectra of multiple stars. As discussed in \S\ref{sec:resmain}, the coadded measurements inherently trend toward the mean of the underlying data, and therefore mask any narrow metal-poor tail in the distribution. In contrast, deriving metallicities for individual stars allows any metal-poor tail to be captured, resulting in a lower overall mean metallicity. To distinguish between this effect and that of the different calibration methods described above would require both the CaT calibration and spectral synthesis to be applied to identical spectra. It is nominally possible to apply the CaT calibration to the coadded spectra used for spectral synthesis; however, this would require assigning a singular V-band magnitude to a coadd of multiple stars that span a range of almost 1.5 magnitudes in brightness. This is neither particularly physically meaningful, nor likely reflective of the coadded spectrum used to perform the measurements given the varying contributions of each individual star in the coadd at any given pixel. Deeper observations of AndXIX, which allow abundances to be derived for individual stars, would eliminate any such magnitude biases in the CaT calibration associated with coaddition and allow for a 1:1 comparison between the abundances derived using the two methods.

\subsection{AndXIX in context}\label{sec:ogalcomp}
\begin{figure*}
	\includegraphics[width=\columnwidth]{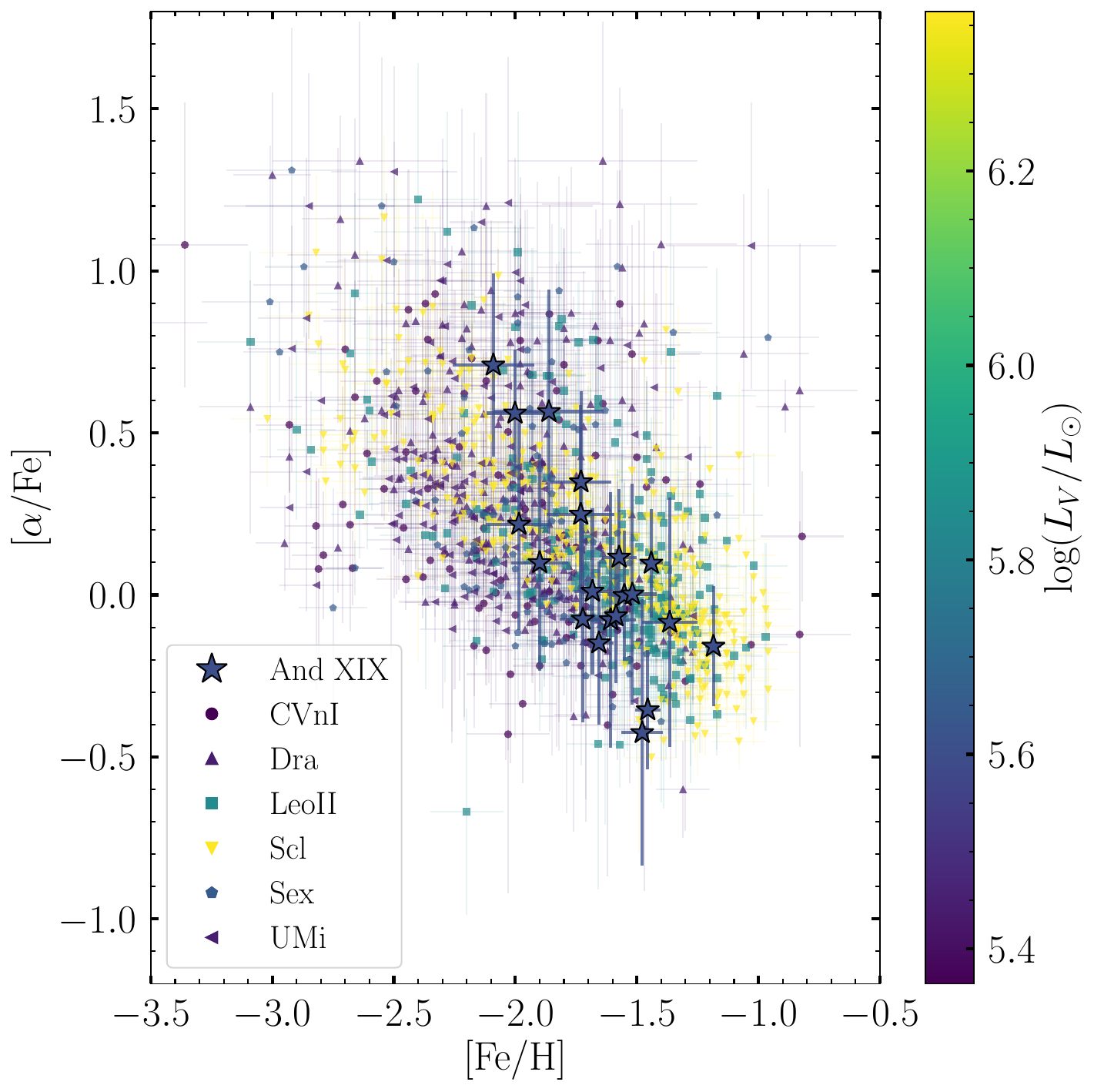}
	\includegraphics[width=\columnwidth]{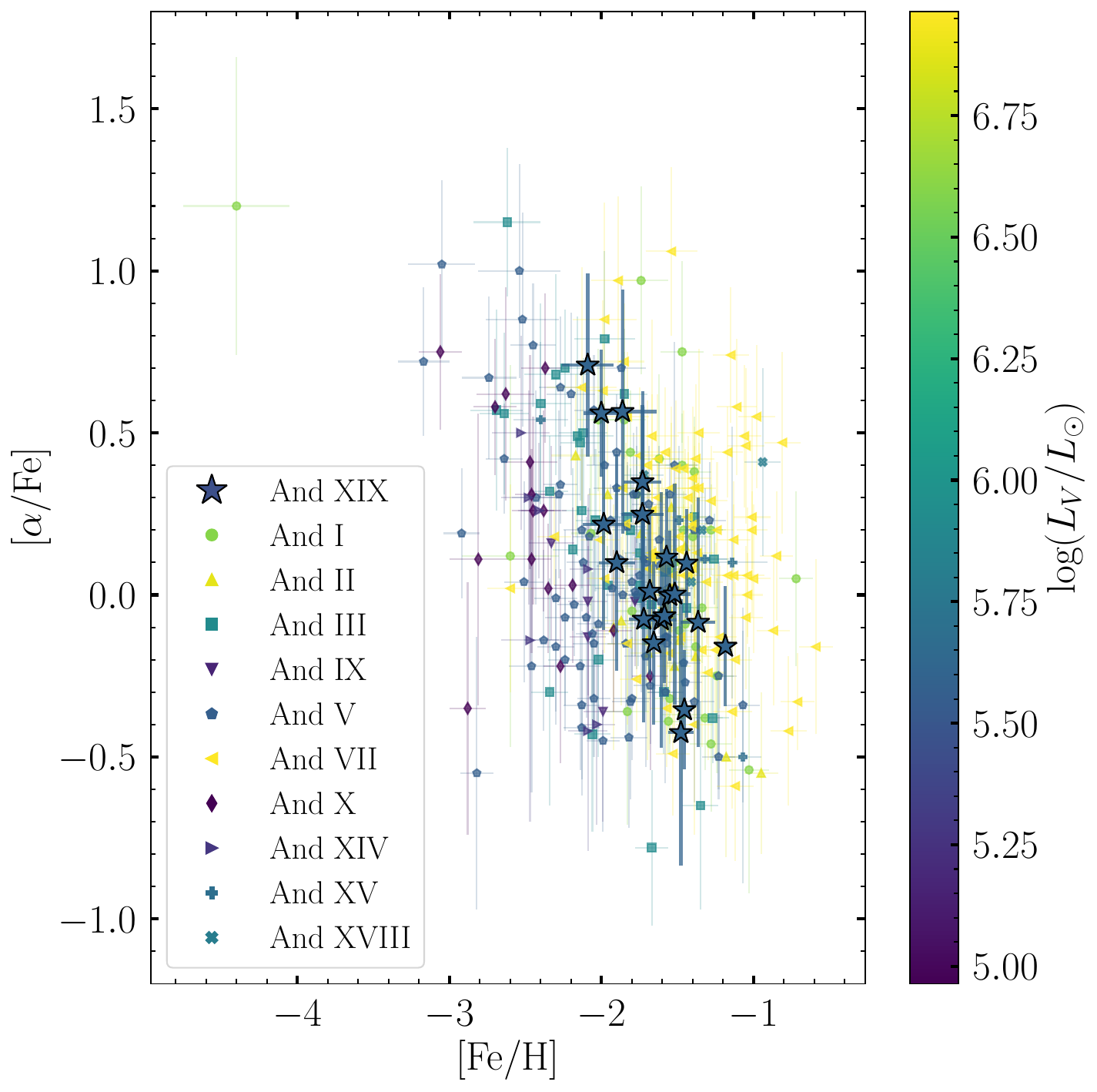}
	\caption{Literature values of [$\alpha$/Fe] vs.\ [Fe/H] for various MW (left) and M31 (right) dSph satellites, colour-coded by log-luminosity (see text for full source list). Our codded AndXIX measurements are indicated by starred points; these have a similar distribution to those of other dSphs.}\label{fig:comp}
\end{figure*}

We compare the abundances we derive for AndXIX to other dSph satellites of the MW and M31 (left and right panels of Fig.~\ref{fig:comp}, respectively) also derived from DEIMOS data using similar techniques. Starred points with black outlines indicate coadded abundances for our full sample; other shapes indicate abundances for other satellites taken from literature, with all points colour-coded by the V-band luminosity of the satellite \citep[all taken from][for consistency]{mcconnachieObservedPropertiesDwarf2012}. All MW satellite abundances are taken from \citet{kirbyMultielementAbundanceMeasurements2010}, and are measured for individual stars. Those abundances are provided for individual $\alpha$-elements (one or more of Mg, Si, Ti, Ca); we calculate an “overall” $\alpha$-element abundance for comparison by taking the average of all available values for a given star, weighted by the inverse square of the individual element uncertainties. For clarity in the figure, we only plot measurements where the [$\alpha$/Fe] uncertainty is $<$0.5~dex. Abundances for all M31 satellites are taken from \citet{wojnoElementalAbundancesM312020}, measured for a combination of individual stars and coadded spectra using the same method as this paper. In addition, abundances obtained from deep spectra of individual stars in select M31 satellites are included where available: we source measurements for And I, III, V, VII and X from \citet{kirbyElementalAbundancesM312020}, taking the bulk [$\alpha$/Fe] even when measurements of individual $\alpha$-elements are available. We do not include individual stellar abundances from \citet{vargasDistributionAlphaElements2014} as \citet{kirbyElementalAbundancesM312020} analyse an overlapping sample of stars with deeper data, and identify a systematic [Fe/H] difference of 0.25~dex between their measurements. Given the method used by \citet{kirbyElementalAbundancesM312020} is more similar to ours than that of \citet{vargasDistributionAlphaElements2014}, we preferentially take the former. %both \citealt{kirbyElementalAbundancesM312020} and \citealt{wojnoElementalAbundancesM312020} note there are no systematic differences in their [$\alpha$/Fe] abundances and those of \citealt{vargasDistributionAlphaElements2014}.

\begin{figure*}
	\includegraphics[width=\columnwidth]{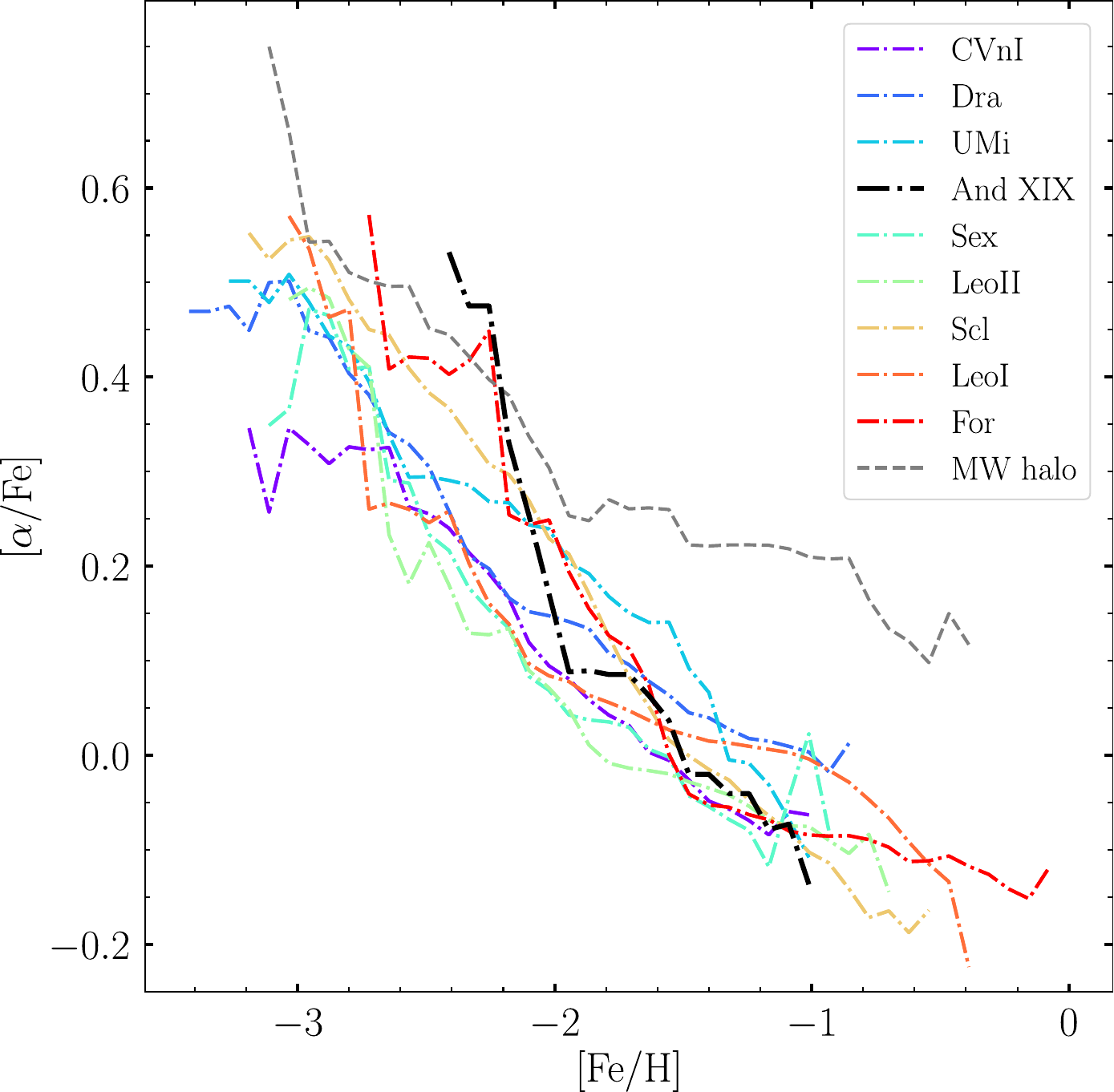}
	\includegraphics[width=\columnwidth]{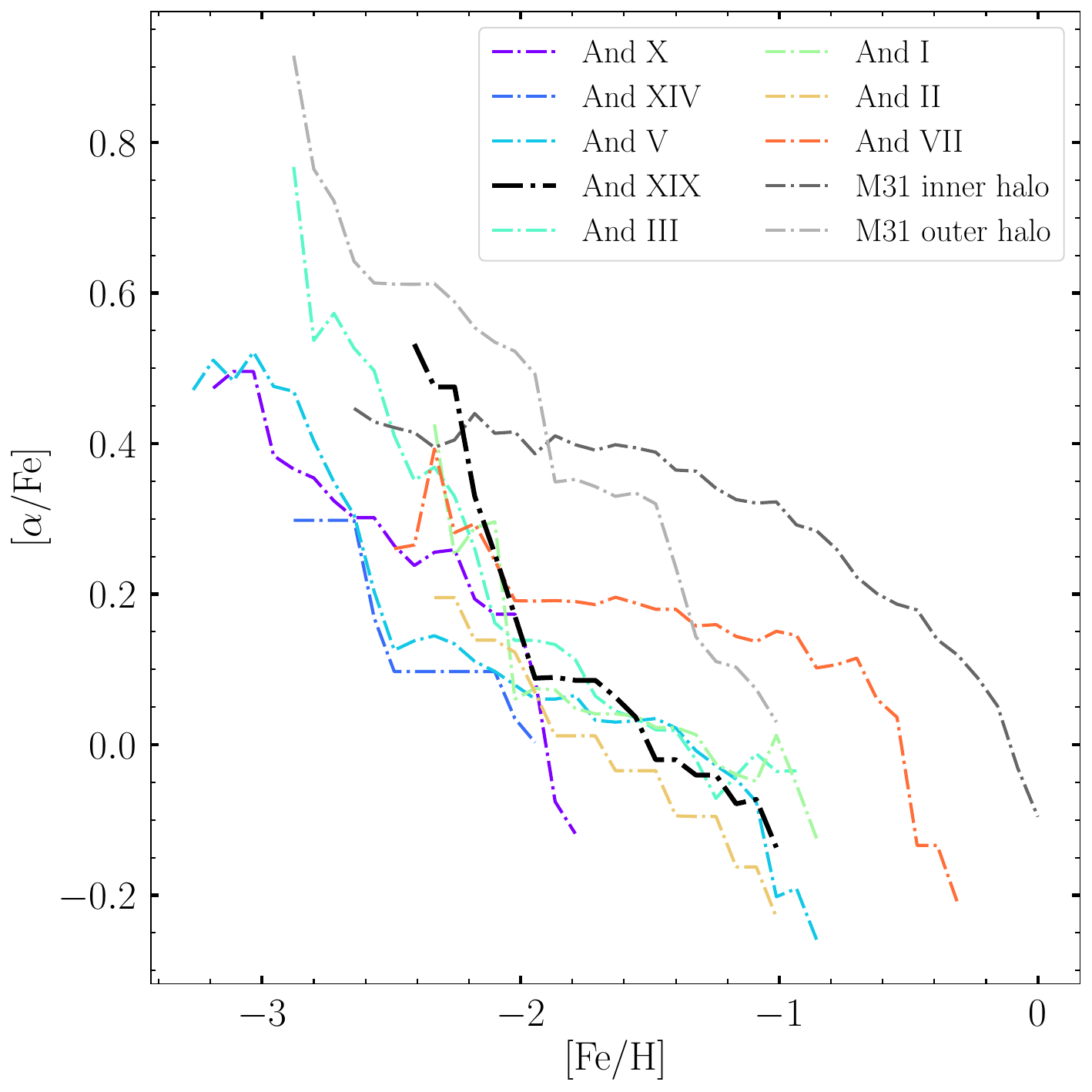}
	\caption{Moving averages of [$\alpha$/Fe] as a function of [Fe/H] for various MW (left) and M31 (right) dSph satellites (see text for full source list). Galaxies are listed from least to most luminous. A solid black line indicates our AndXIX measurements. Dashed grey lines indicate average values for the MW and M31 (inner and outer, defined as within and beyond 30~kpc, respectively) halos in their respective panels. AndXIX displays a very similar trend to that seen in most other dSphs.} \label{fig:avcomp}
\end{figure*}

The abundances we measure for AndXIX are comparable with those of similar-luminosity satellites, both in terms of overall [$\alpha$/Fe] and [Fe/H] abundances, and their associated dispersions. We confirm this by performing 2D K--S tests, comparing the [$\alpha$/Fe]-[Fe/H] distribution of AndXIX to combined distributions of the MW and other M31 satellites; in both cases, there are no significant differences in the abundance distributions. In Fig.~\ref{fig:comp}, this is clearest in the comparison with other M31 satellites, but is true also in the comparison to MW satellites; here, the more numerous measurements for satellites of higher and lower luminosities (only Sextans has a very similar luminosity to AndXIX) wash out the trend. To make this clearer, in Fig.~\ref{fig:avcomp}, we present the moving average of the [$\alpha$/Fe] abundances as a function of [Fe/H] for the MW (left) and M31 (right) satellites respectively. This is calculated by taking the average of [$\alpha$/Fe] measurements, weighted by the inverse square of the measurement uncertainties, in moving windows of 0.5~dex in [Fe/H]; we exclude bins where $<5$ [$\alpha$/Fe] measurements contribute to the average. This means some M31 satellites with very few abundance measurements are excluded from the plot entirely; as are the tails of the [Fe/H] distribution that include very few measurements. In addition, in these plots, we show in grey the average [$\alpha$/Fe] vs.\ [Fe/H] trend for the MW halo \citep{mckinnonHALO7DIIIChemical2023} and the inner \citep[$<30$~kpc:][]{escalaElementalAbundancesM312020a,wojnoElementalAbundancesM312023} and outer \citep[$>30$~kpc:][]{wojnoElementalAbundancesM312023} halo of M31. We choose these reference samples as they are all derived using very similar methods on similar DEIMOS data, which should minimise any potential systematic differences. 

Qualitatively, the decrease in [$\alpha$/Fe] as a function of [Fe/H] for AndXIX is very similar to that seen in other dSphs, and the outer halo of M31 -- which \citealt{wojnoElementalAbundancesM312023} suggest is likely comprised of debris accreted from several such small satellites. While it appears that the M31 satellites And V \& X have significantly lower [$\alpha$/Fe] abundances than their MW counterparts at the metal-rich end of their distributions, this appears likely to be at least partially an artifact due to the low number of stars in the surrounding bins for these two galaxies, compared to that of their MW counterparts at the same metallicity. This means that a handful of low-$\alpha$ stars in this bin, well within the overall abundance scatter of the galaxy, can have an outsized effect on the running mean reported in Fig.~\ref{fig:avcomp}. We also note there is no clear evidence of metal-poor plateaus for any galaxies in Fig.~\ref{fig:avcomp}. As discussed in \S\ref{sec:resmain}, this is due to the underlying limitations of the chosen comparison samples (which are shared with our AndXIX data) -- higher-resolution studies of several MW satellites do find high [$\alpha$/Fe] plateaus at very low metallicities \citep[e.g.][]{lemasleVLTFLAMESSpectroscopy2014,hillVLTFLAMESHighresolution2019,thelerChemicalEvolutionDwarf2020}.

The slope of the [$\alpha$/Fe]-[Fe/H] trend, as seen in Fig.~\ref{fig:avcomp} is linked to the chemical evolution history and mass of a system. The shallow gravitational potential well of low-mass systems facilitates gas loss due to stellar feedback \citep{dekelOriginDwarfGalaxies1986}, which not only suppresses the increase of [Fe/H] in the system, but also results in a lower overall SFR, allowing the rate of Type Ia SN to dominate that of core collapse SN and producing a steeper [$\alpha$/Fe] slope \citep{kirbyElementalAbundancesM312020}. In contrast, more massive systems better retain their (enriched) gas, allowing for a higher rate of star formation over an extended period. This not only results in a comparatively higher rate of Type II supernovae, but also increases the overall [Fe/H] of the system, both of which produce a shallower [$\alpha$/Fe] gradient \citep{kirbyElementalAbundancesM312020}. 

This is clearly seen in the inner halo of M31, which is likely comprised of debris from more massive satellites than those which survive today \citep{escalaElementalAbundancesM312020,kirbyElementalAbundancesM312020}, and which has a correspondingly shallower slope in Fig.~\ref{fig:avcomp}. The slope of the MW halo in Fig.~\ref{fig:avcomp} is also somewhat shallower, particularly at the metal-rich end, but is comparatively steep at the metal-poor end. This is because the \citet{mckinnonHALO7DIIIChemical2023} sample includes MW halo stars with a variety of origins. The metal-rich end is dominated by `in-situ' halo stars heated from the MW disk and debris from Gaia-Sausage-Enceladus -- which, being relatively massive, has a shallow [$\alpha$/Fe] gradient \citep[e.g.][]{hasselquistAPOGEEChemicalAbundance2021} -- while the metal-poor end is comprised of debris from smaller satellites, with accordingly shallower [$\alpha$/Fe] gradients \citep[e.g.][]{naiduEvidenceH3Survey2020}.

It is also worth noting the diversity of [$\alpha$/Fe]-[Fe/H] slopes in M31 satellites compared to those in the MW. While the lower-mass MW satellites (i.e. Sextans and below) largely overlap in this space -- with this similarly being observed in higher-resolution studies \citep[e.g.][]{fernandesComparativeAnalysisChemical2023} -- the [$\alpha$/Fe]-[Fe/H] locus of M31 satellites varies significantly. To quantify this, we measure the slope d[$\alpha$/Fe]/d[Fe/H] using the method described in section 7 of \citet{wojnoElementalAbundancesM312020}. That work shows such a slope derived from coadded measurements accurately reflects that derived from abundance measurements of individual stars. In particular, we parameterize the trend with an angle ($\theta$), and the perpendicular distance of the line from the origin ($b_\perp$) per Eq.~\ref{eq:slope}.

\begin{equation}\label{eq:slope}
	[\alpha/\text{Fe}] = [\text{Fe/H}]\tan(\theta) + \frac{b_\perp}{\cos(\theta)}
\end{equation}

This avoids the preference for shallow slopes when the d[$\alpha$/Fe]/d[Fe/H] slope is fit directly with a flat prior \citep{hoggDataAnalysisRecipes2010}. We sample the posterior distribution of the parameters (both with flat priors) using the Markov chain Monte Carlo ensemble sampler \textsc{emcee} \citep{foreman-mackeyEmceeMCMCHammer2013}, taking the 50th percentile of the resulting distribution as our measured value, and the 16th and 84th percentiles as the associated uncertainties. We then back-calculate the associated slope as d[$\alpha$/Fe]/d[Fe/H]$=\tan(\theta)$. A slope of $-1$ would imply constant [$\alpha$/H], i.e., enrichment solely by Type Ia SN with no contribution from core collapse supernovae \citep{kirbyElementalAbundancesM312020}. 

\begin{figure}
	\includegraphics[width=\columnwidth]{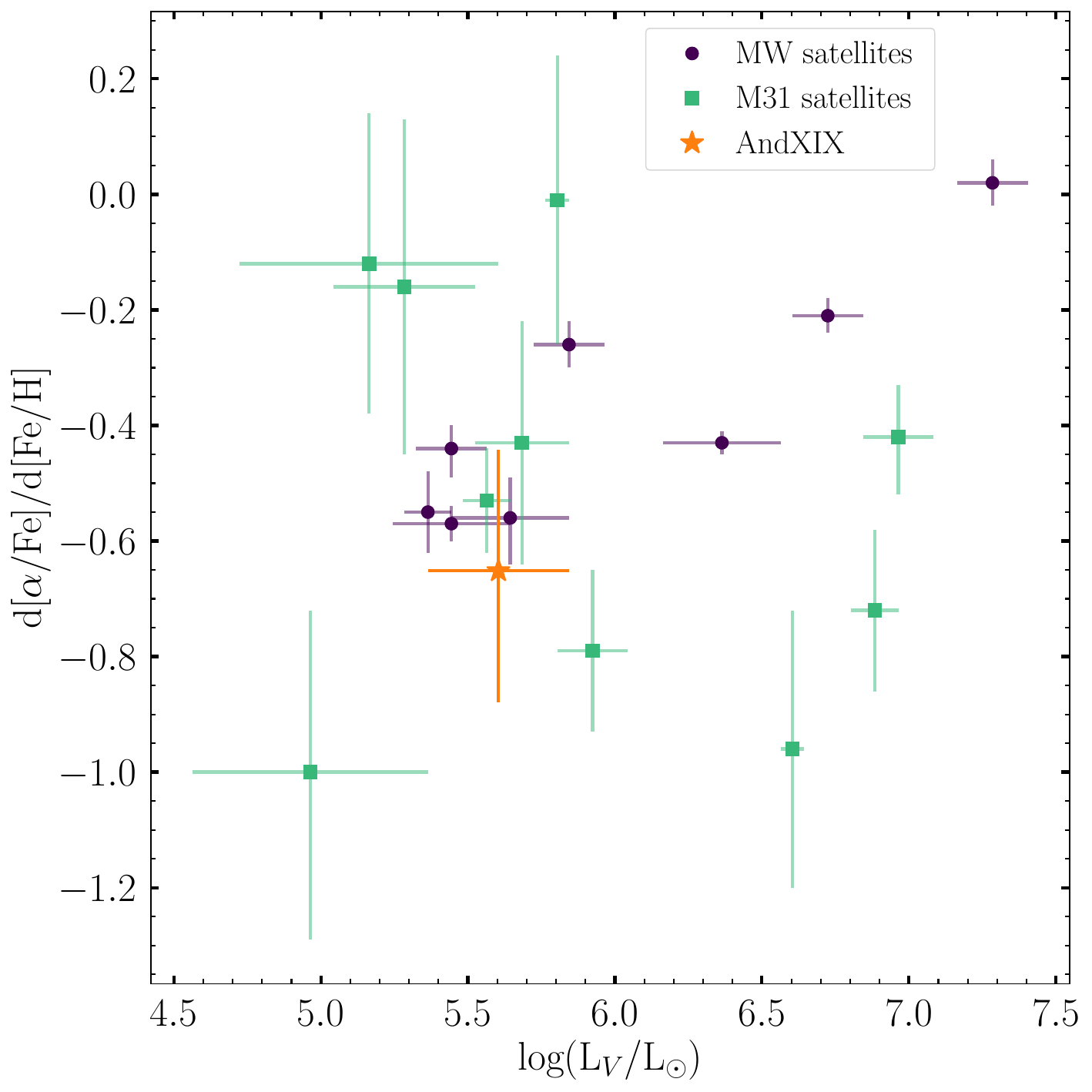}
	\caption{d[$\alpha$/Fe]/d[Fe/H] slope as a function of log-luminosity for Local Group dSphs (see text for full source list). Steeper (i.e.\ more negative) slopes indicate a stronger contribution of Type Ia SN to the chemical evolution of the galaxy. Purple circles indicate MW dSphs and green squares indicate M31 dSphs; the starred orange point represents our measurement for AndXIX\@.} \label{fig:slope}
\end{figure}

We measure a d[$\alpha$/Fe]/d[Fe/H] slope of $-0.65^{+0.21}_{-0.23}$ for AndXIX, and compare this to literature measurements of this slope reported for other dSphs using the same method in Fig.~\ref{fig:slope}. Measurements for MW dSphs are taken from \citet{kirbyElementalAbundancesM312020}. For M31 satellites, we preferentially take measurements from \citet{kirbyElementalAbundancesM312020}, as these are derived from larger samples of individual stellar abundances; the remaining M31 satellites (And II, IX, XIV, XV, and XVIII) are sourced from \citet{wojnoElementalAbundancesM312020}. 

The [$\alpha$/Fe]-[Fe/H] gradient in AndXIX is relatively steep, and is consistent with that of similar-luminosity dSphs, such as Sextans. This implies a similar chemical evolution history for these galaxies, which aligns with the similar SFHs observed. For example, the SFH of Sextans indicates that it formed the majority of its stars in the first 1-2~Gyr after the big bang, with minimal to no SF more recently than 11~Gyr ago \citep{leeStarFormationHistory2009,bettinelliStarFormationHistory2018}; this is very similar to the SFH of AndXIX, which also experienced the majority of its star formation prior to reionization, and has no star formation within the last 8-10~Gyr (\citetalias{Col22}). In general, however, there is significantly more scatter in this plane for M31 satellites compared to MW satellites, which clearly trend toward shallower slopes with increasing luminosity. This is likely linked to global differences in the SFHs, and therefore chemical evolution, of M31 and MW satellites. In particular, \citet{weiszComparingQuenchingTimes2019} find that the time at which star formation is quenched -- with more recent quenching times resulting in higher metallicities and shallower [$\alpha$/Fe]-[Fe/H] slopes -- is effectively independent of luminosity for M31 satellites, but is strongly correlated with luminosity for MW satellites \citep[e.g.][]{brownQUENCHINGULTRAFAINTDWARF2014, weiszSTARFORMATIONHISTORIES2015}.

\section{Discussion: the origin of AndXIX}\label{sec:origins}
Multiple different scenarios have been proposed to explain the diffuse appearance of AndXIX; here, we discuss what our abundance measurements, in conjunction with its other observed properties, imply for its formation. 

\subsection{Tidal interactions}\label{sec:tides}
In this scenario, AndXIX experienced a strong tidal interaction with a much larger system (the most likely candidate being M31), which resulted in its dark and baryonic matter being redistributed outward -- and/or stripped -- to produce the diffuse system observed today. Even before considering abundances, AndXIX possesses numerous characteristics suggestive of it having experienced tidal interactions. These include distorted outer isophotes \citep{mcconnachieTrioNewLocal2008}; a mild ($-2.1\pm1.7$~km~s$^{-1}$~kpc$^{-1}$) velocity gradient and spatially varying velocity dispersion \citepalias{C19}; and its vicinity to other low-surface-density features. In particular, it is near what \citet{ibataHauntedHalosAndromeda2007} first identify as diffuse structure aligned with the major axis of M31, and which extended data in \citet{mcconnachieTrioNewLocal2008,mcconnachieLargescaleStructureHalo2018} show as a broader complex of substructure which extends to the northeast of AndXIX, though no clear association between AndXIX and these diffuse structures has been proven. In addition, the nominal association of AndXIX with a nearby streamlike feature to the southwest is debated given an $\sim$0.4~dex metallicity difference between the systems \citepalias{C19}.

\begin{figure}
	\includegraphics[width=\columnwidth]{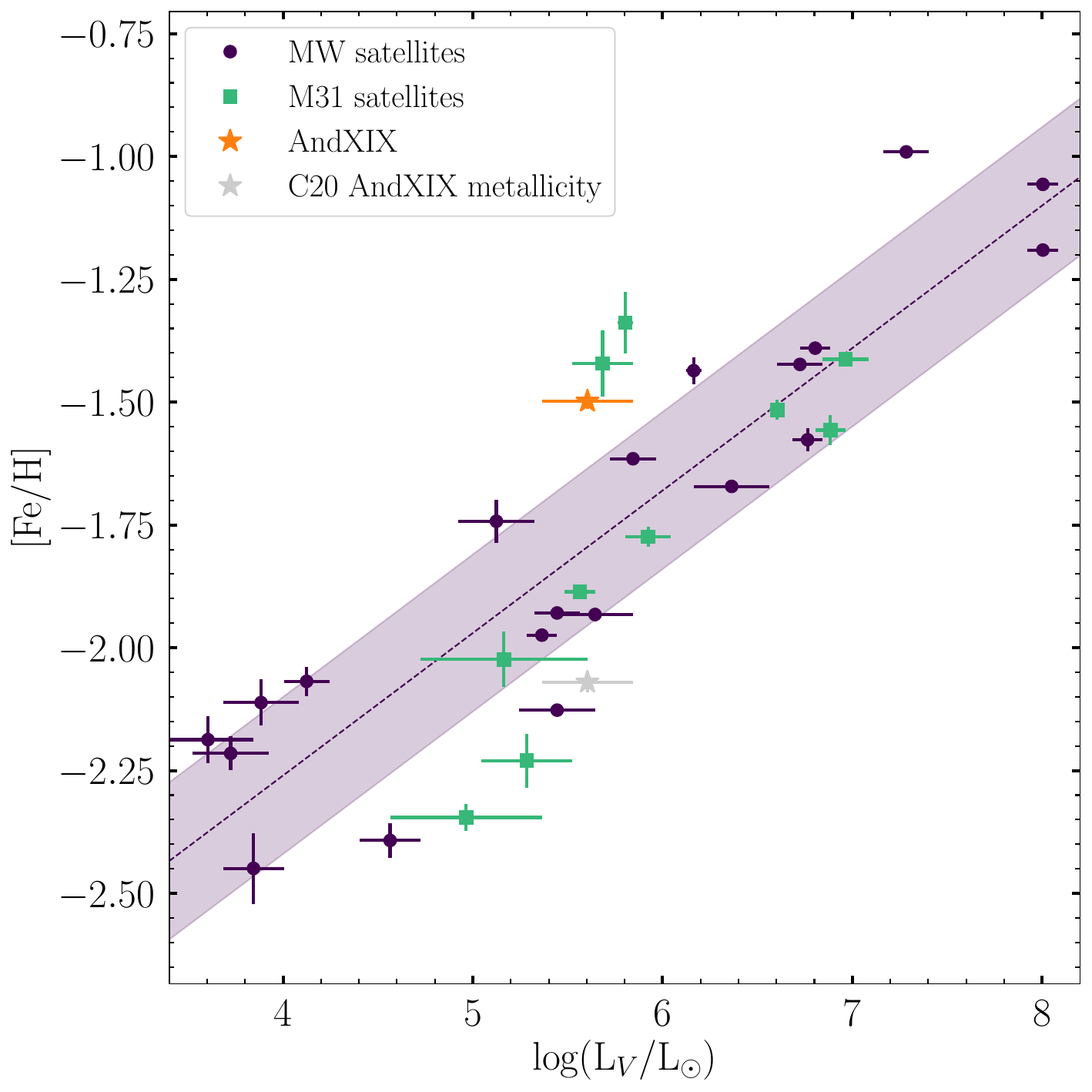}
	\caption{Luminosity-metallicity relation for Local Group dSphs. The purple dashed line and shaded regions represent the mass-metallicity relation and its associated root-mean-square scatter of 0.16~dex from \protect\citet{kirbyUniversalStellarMassStellar2013}. Purple circles indicate MW dSphs and green squares indicate M31 dSphs; points without [Fe/H] error bars have sufficiently small ranges (due to being averages of many data points) that these are not visible. The grey starred point represents the previous CaT-based AndXIX metallicity from \citetalias{C19}; our new measurement (orange starred point) is $\sim0.6$~dex more metal-rich (see \S\ref{sec:c19rcomp}), placing it $\sim2\sigma$ above the mass-metallicity relation. } \label{fig:massmet}
\end{figure}

If AndXIX has experienced significant tidal stripping as a result of interactions, we expect it to have a mean metallicity higher than that predicted by the present-day stellar mass-metallicity relation for dSphs \citep{kirbyUniversalStellarMassStellar2013}. The previous measurement of AndXIX’s metallicity by \citetalias{C19} placed it approximately 2$\sigma$ below this relation, making this unlikely; however, as per \S\ref{sec:c19rcomp}, our mean metallicity is $\sim$0.6~dex higher. We thus compare our updated metallicity to the \citet{kirbyUniversalStellarMassStellar2013} relation in Fig.~\ref{fig:massmet}. The orange star represents our new measurement, calculated as the average of all coadd [Fe/H] measurements, weighted by the inverse square of their uncertainties; the grey star shows the previous measurement from \citetalias{C19}. Our new [Fe/H] measurement now places AndXIX approximately 2$\sigma$ above the mass-metallicity relation. 

If AndXIX was originally on the mass-metallicity relation, this implies it had a luminosity of $\sim4.5\times10^6$~L$_\odot$; assuming a stellar mass-to-light ratio of 1.6~M$_\odot$/L$_\odot$ -- the average value for dSphs measured by \citet{wooScalingRelationsFundamental2008} -- implies an initial stellar mass of $\sim7.2\times10^6$~M$_\odot$. This is a factor of ten higher than AndXIX’s current stellar mass assuming the same M/L ratio ($\sim6.4\times10^5$~M$_\odot$) -- requiring a loss of $\sim90$\% of its initial stellar mass. Even if AndXIX were originally at the upper envelope of the $1\sigma$ scatter within the mass-metallicity relation, this still implies a minimum loss of $\sim65$\% of its initial stellar mass. Strong tidal interactions are required for such significant stripping. 

Simulations by \citet{penarrubiaColdDarkMatter2008,penarrubiaTidalEvolutionLocal2008} support this picture. They find a loss of 90\% of a satellite galaxy's stellar mass can maintain (or increase) its half-light radius -- consistent with the large radius of AndXIX -- while leaving its internal kinematics largely unchanged. In addition, they find tidal stripping tends to increase the \textit{total} mass-to-light ratio; this is consistent with unusually large total M/L ratio measured for AndXIX by \citetalias{C19}.  

If AndXIX was originally $\sim10\times$ more massive than its current luminosity suggests, this might allow it to initially retain more gas than satellites with similar present-day luminosities, and thus reach a higher metallicity. This would explain the mild horizontal shift seen in Fig.~\ref{fig:avcomp} of its [$\alpha$/Fe]-[Fe/H] trend towards higher [Fe/H] values relative to the currently similar-mass satellites Sextans and And V, closer to that of e.g.\ And I and Leo I, which have stellar masses closer to that of AndXIX’s nominal original mass assuming it began on the mass-metallicity relation. This is particularly clear at [$\alpha$/Fe] values $>0.1$ (i.e.\ [Fe/H] values $<-2$), where AndXIX is up to $0.5$~dex higher in metallicity than the satellites most similar to its current luminosity.  

The steep d[$\alpha$/Fe]/d[Fe/H] gradient we observe for AndXIX is not inconsistent with this picture -- an early tidal interaction that strips gas from AndXIX could mimic the effect of gas loss via internal mechanisms in lower-mass galaxies, producing a similarly steep [$\alpha$/Fe]-[Fe/H] gradient. The SFH of AndXIX, which indicates it quenched $\sim$10~Gyr ago and has not experienced any recent star formation (\citetalias{Col22}), is consistent with such a picture. 

Overall, we consider that tidal interactions are a likely candidate for the formation of AndXIX\@. If those tidal interactions were with M31, this may indicate AndXIX is on a highly radial orbit, since simulations suggest tidal interaction mechanisms are most effective when satellites are on strongly radial orbits and thus experience close interactions \citep[e.g.][]{maccioCreatingGalaxyLacking2020,morenoGalaxiesLackingDark2022}. This is the case for the similarly diffuse Antlia II, which has a close orbital pericentre of $38.6^{+7.2}_{-5.8}$~kpc around the MW, and is thought to be the product of tidal interactions \citep{jiKinematicsAntliaCrater2021}. Proper motion measurements are necessary to accurately model the orbit of AndXIX and confirm if tidal interactions with M31 are a plausible formation mechanism. If the proper motions of AndXIX are aligned with its velocity gradient (which, per \citetalias{C19}, is aligned with the major axis of the galaxy), this would be further indicative of tidal interactions as its origin; a similar alignment of these three features is observed in Antlia II \citep{jiKinematicsAntliaCrater2021}. 

\subsubsection{Orbital modelling}\label{sec:orbits}
Motivated by the above scenario, we perform a preliminary test of whether AndXIX is likely to have experienced close tidal interactions with M31. We use the framework described in \citet{kvasovaKinematicsMetallicityDwarf2024} (see their Appendices B and C for details), and investigate the effects of different tangential velocities ($v_\tau$) of AndXIX relative to M31 on its possible orbit. For simplicity, we consider only the two-body problem (AndXIX-M31), assuming a point-like, static potential for both galaxies. Table~\ref{tab:modelpars} presents the parameters we use for the models. 

{\setlength{\tabcolsep}{2pt} 
	\begin{deluxetable}{lrl} \label{tab:modelpars}
		\tablewidth{\columnwidth}
		\tabletypesize{\footnotesize}
		%% This is the title of the table.
		\tablecaption{Masses, positions, and kinematics used for AndXIX orbital modelling.}
		%% The \tablehead gives provides the column headers.  It is currently set up so that the column labels are on the top line and the units surrounded by ()s are in the bottom line. 
		\tablehead{\colhead{Parameter} & \colhead{Value} & \colhead{Source}} 
		%% All data must appear between the \startdata and \enddata commands
		\startdata
		M31 mass & $1.2\times10^{12}$~M$_\odot$ & \citet{vandermarelM31VELOCITYVECTOR2012a} \\
		AndXIX mass & $1.1\times10^8$~M$_\odot$ & \citetalias{C19} \\
		MW-M31 distance & 770~kpc & \citet{karachentsevMassesLocalGroup2006} \\
		MW-AndXIX distance & 821~kpc & \citet{connBayesianApproachLocating2012} \\
		AndXIX-M31 distance & 115~kpc & \citet{connBayesianApproachLocating2012} \\
		$V_{\text{LOS}}$, M31 & $-301$~km~s$^{-1}$ & \citet{karachentsevMassesLocalGroup2006} \\
		$V_{\text{LOS}}$, AndXIX & $-110$~km~s$^{-1}$ & \citetalias{C19} \\
		\enddata
\end{deluxetable} }

\begin{figure*}
	\centering
	\includegraphics[width=\textwidth]{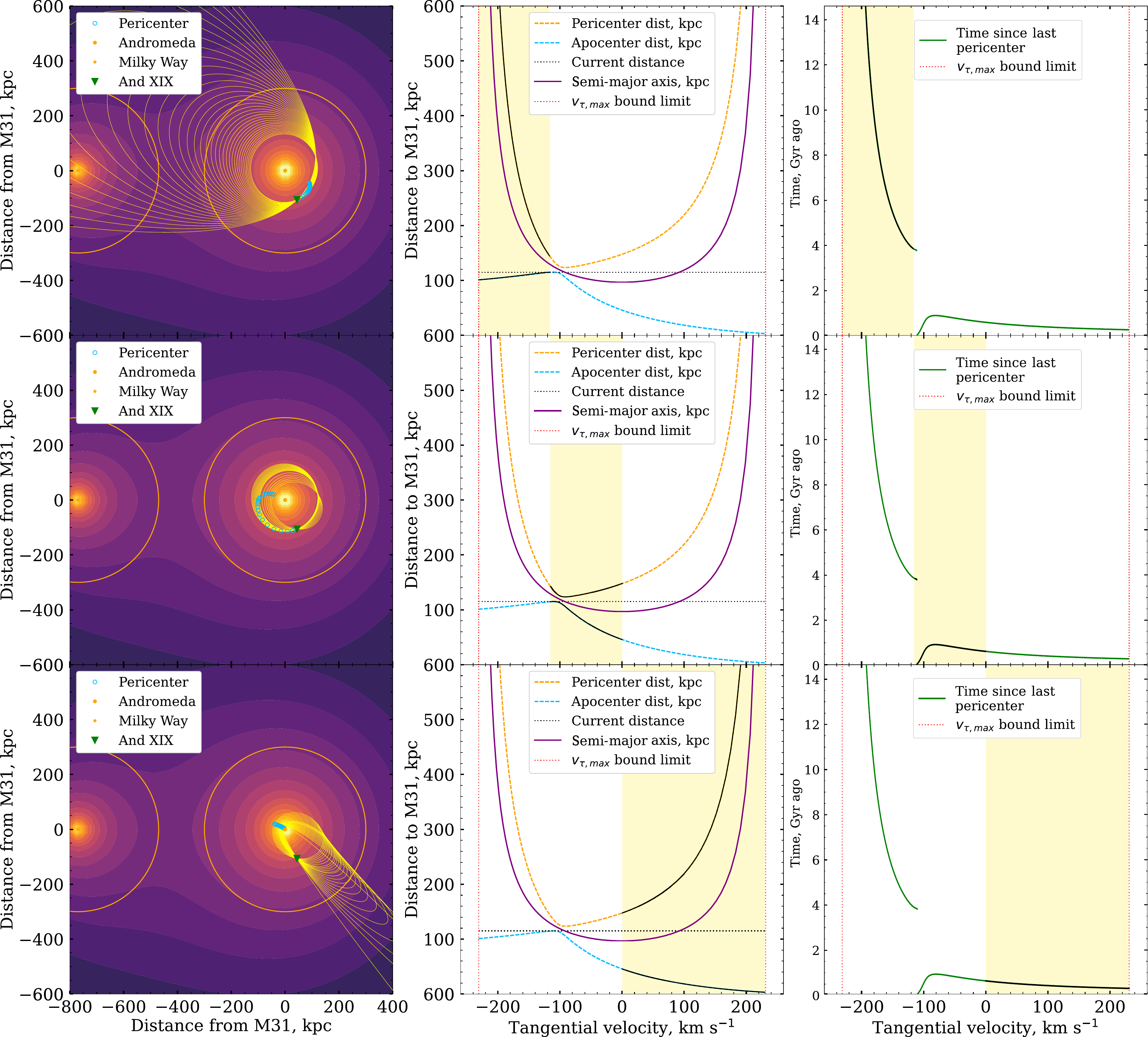}
	\caption{\textit{Left column:} Possible orbits for AndXIX around M31 in a two-body framework for different relative tangential velocities $v_\tau$; the tangential velocity is positive along the negative y-axis. Each yellow line shows a singular orbit (i.e.\ apocentre to apocentre) for a given value of $v_\tau$ which passes through AndXIX's current position, sampled from the yellow shaded region in the right column. The background contour plot illustrates the equipotential surfaces for the MW (for reference only) and M31, with orange circles indicating their virial surfaces (at 300~kpc). Green triangles indicate AndXIX's current position. Blue circles indicate the pericentre associated with each orbit. Different rows indicate different ranges of orbits associated with different ranges of tangential velocities: $v_\tau<109$~km~s$^{-1}$ (top), $-109<v_\tau$~km~s$^{-1}<0$ (middle), and $v_\tau>0$~km~s$^{-1}$ (bottom). \textit{Middle column:} orbital properties of AndXIX associated with different tangential velocities. The horizontal dotted black line indicates AndXIX's current distance from M31; the vertical red dotted lines correspond to $|v_\tau|=231$~km~s$^{-1}$ (the limit between which AndXIX is bound to M31). Yellow dashed and dotted lines indicate the pericentric and apocentric distances from M31; these are shown as solid black lines for the range of tangential velocities plotted in the corresponding row. The purple line indicates the semi-major axis of the orbit. \textit{Right column:} time since AndXIX's last pericentric passage as a function of tangential velocity (green), shown as a solid black line for the range of tangential velocities plotted in the corresponding row. Both strongly radial and relatively circular orbits are possible for AndXIX; these can be grouped into three qualitatively different scenarios, differentiated by the assumed tangential velocity (\S\ref{sec:orbits}).}
	\label{fig:body2}
\end{figure*}

Fig.~\ref{fig:body2} shows different possible orbits of AndXIX around M31 in the two-body scenario; the orbits are projected onto an X-Y plane along the line of sight from the observer to M31 such that a positive radial velocity is to the right (away from the MW), and a positive tangential velocity $v_\tau$ is downwards (i.e. along the negative y-axis). Both the radial and tangential velocities are relative to M31's systemic motion. The system is closed (i.e.\ AndXIX is bound to M31) while $|v_\tau|<231.3$~km~s$^{-1}$; AndXIX is currently at pericentre if $v_\tau=-109.7$~km~s$^{-1}$.

We find there are broadly three qualitatively different families of orbits for AndXIX that can reproduce the observed data. The first are those where it has a tangential velocity $v_\tau<-109.7$~km~s$^{-1}$ (top row of Fig.~\ref{fig:body2}); in this scenario, it has experienced at least one previous pericentric passage around M31 on the order of several Gyr ago (and has not yet reached pericentre again). These orbits all have fairly distant pericentres, comparable to AndXIX's current distance from M31 ($\sim115$~kpc). This aligns with the scenario predicted for AndXIX by \citet{watkinsCensusOrbitalProperties2013}, who suggest an orbital pericentre for AndXIX of $112\pm12$~km~s$^{-1}$ based on the timing argument. %While this is sufficiently close that ram pressure or tidal stripping can remove gas from AndXIX in such orbits, consistent with its ancient quenching \citepalias{Col22}, 
However, these orbits are not the strongly radial ones necessary to substantially tidally disrupt the stellar component of AndXIX\@.

The second category of orbits correspond to $-110\lesssim v_\tau\lesssim0$~km~s$^{-1}$ (second row of Fig.~\ref{fig:body2}). In this family of orbits, AndXIX is on a mostly circular orbit, with broadly similar peri- and apocentric distances ($\sim50-150$~kpc), and has recently experienced a pericentric passage around M31. These are again too distant to substantially tidally perturb the central regions of AndXIX\@. 

The third category of orbits corresponds to $v_\tau\gtrsim0$~km~s$^{-1}$ (bottom row of Fig.~\ref{fig:body2}). These orbits have fairly long periods, with AndXIX only experiencing a pericentric passage relatively recently (within the past $\sim$Gyr). They are, however strongly radial orbits, with pericentric distances within $\sim50$~kpc of M31: certainly strong enough to tidally disturb AndXIX and produce the diffuse galaxy we see today. 

Among all the range of orbits discussed above, the MW has relatively little effect. While some orbits appear to pass close to the MW in the top row of Fig.~\ref{fig:body2}, these both a) do not account for the increased distance between the MW and M31 in the past (since this is a two-body-only model), and b) in some cases correspond to sufficiently long-period orbits that such passages would actually occur in the future, as the left column of Fig.~\ref{fig:body2} plots one full orbital period, regardless of whether sufficient time has passed for AndXIX to actually complete such an orbit. 

The above modelling suggests that there are several reasonable orbital solutions -- with $v_\tau\gtrsim0$~km~s$^{-1}$ -- where AndXIX has passed close enough to M31 to have experienced strong tidal interactions within the past $\sim$Gyr. However, as sufficient time must have passed since a close interaction for AndXIX to expand and return to virial equilibrium to produce its current observed properties, this suggests orbits where this interaction is very recent (corresponding to the largest $v_\tau$) are implausible. Our models also suggest it is unlikely that a singular interaction has led to both the quenching and tidal expansion of AndXIX, due to the very different timescales of these events ($\gtrsim8$~Gyr ago and $\lesssim1$~Gyr ago respectively).

These are, however, very simple models. More sophisticated modelling -- beyond the scope of this paper, but which includes e.g.\ dynamical friction (not present in these models), a more realistic shape for M31's gravitational potential (rather than the pointlike one used in these models), the growth of M31's dark matter halo over time (fixed at its current mass in these models, which is a particularly bad assumption at ancient times concurrent with AndXIX's SF quenching), and the effect of massive satellites like M33 and the progenitor of M31's Giant Stellar Stream on M31's gravitational potential, which can subsequently affect the orbits of its satellites \citep[similarly to how to the LMC affects the gravitational potential of the MW and its satellites: e.g.][]{gomezItMovesDangers2015} -- is critical in order to test whether there are truly orbits which can naturally explain all of AndXIX's properties. 

Given all of the above uncertainties, it is quite possible AndXIX is not on a long-term bound orbit around M31, and therefore has not closely interacted with it in the past. However, even if future proper motion measurements reveal AndXIX is on an orbit such that it has not closely interacted with M31, this may not rule out tidal interactions as a formation mechanism. It is also possible AndXIX could have tidally interacted with another galaxy $\sim8-10$~Gyr ago \citep[e.g.\ the progenitor of M31's last, likely major, merger:][]{hammerBillionYearOld2018}. More detailed simulations of the entire M31 system would be required to confirm or deny such a scenario. 

\subsection{Bursty star formation}\label{sec:burstysf}
In this scenario, feedback from vigorous star formation in AndXIX deposited energy back into its dark and baryonic matter, causing it to expand into the diffuse system seen today \citep[e.g.][]{dicintioNIHAOXIFormation2017, readDarkMatterHeats2019}. However, as discussed in \citetalias{Col22}, the purely ancient SFH of AndXIX does not align with this potential formation mechanism; \citet{readDarkMatterHeats2019} find extended star formation, to within the last $\sim6$~Gyr, is necessary to significantly impact the central DM density of a galaxy, inconsistent with AndXIX’s quenching $\sim10$~Gyr ago. In addition, \citet{chanOriginUltraDiffuse2018} find the effects of stellar feedback on dark and baryonic matter are strongest in galaxies with stellar masses $M_*\sim10^8$~M$_\odot$. This is three orders of magnitude larger than AndXIX’s current stellar mass ($\sim6\times10^5$~M$_\odot$). Even if AndXIX has undergone significant tidal stripping, as described in \S\ref{sec:tides}, its nominal original mass based on the current-day mass-metallicity relation -- $\sim7\times10^6$~M$_\odot$ -- is still well below the regime where bursty star formation is thought to have a significant effect. The steep [$\alpha$/Fe] gradient we measure for AndXIX is also inconsistent with extended bursty star formation as a formation mechanism. In this scenario, shallow [$\alpha$/Fe] slopes are expected, due initially to the dominance of Type II SN during the vigorous SF bursts, and subsequently to the reaching of equilibria between Type II and Type Ia SN over the extended duration of the repeated bursts \citep{revazDynamicalChemicalEvolution2009}. We can therefore confidently rule out extended, bursty SF as the formation mechanism of AndXIX\@. 

\subsection{Mergers}\label{sec:mergers}
There are several different merger-related channels which could nominally result in the formation of AndXIX, each of which will imprint different signatures on its properties. 

In the high-velocity collision scenario, two relatively gas-rich dwarf galaxies collide $\gtrsim10$~Gyr ago at a relatively high velocity ($\sim300$~km~s$^{-1}$); in an event similar to that observed in the Bullet Cluster \citep{cloweDirectEmpiricalProof2006}, the dark matter halos of the galaxies pass through each other, while their dissipative baryons interact to form a new, diffuse galaxy with relatively low dark matter density \citep[e.g.][]{silkUltradiffuseGalaxiesDark2019,shinDarkMatterDeficient2020,leeDarkMatterDeficient2021,otakiFrequencyDarkMatter2023}. However, the vigorous star formation (during which Type II SN should dominate over Type Ia SN) which occurs as a result of the collision to form the new galaxy should produce a relatively shallow [$\alpha$/Fe] gradient as a function of [Fe/H] \citep{silkUltradiffuseGalaxiesDark2019}, unlike the steep decline we observe for AndXIX\@.

In the late dry merger scenario, several mergers between small, gas-poor satellites built AndXIX into an inherently diffuse galaxy \citep{reyEDGEOriginScatter2019}. Since these small systems are thought to be quenched by reionization, the resulting galaxy should be very metal-poor, and -- since there should be insufficient time for many Type Ia SN to have occurred pre-reionization -- will have a relatively shallow [$\alpha$/Fe] gradient as a function of [Fe/H]. This does not align with the metallicity we measure for AndXIX, which is $\sim2\sigma$ \textit{above} the mass-metallicity relation, nor the steep decline in [$\alpha$/Fe] we observe. Truly dry mergers also cannot easily explain the second post-reionization burst of SF in AndXIX seen by \citetalias{Col22}.

In the tidal dwarf scenario, baryonic material ejected during the interaction or merger of two massive, gas-rich galaxies becomes self-gravitating, producing a diffuse, effectively dark-matter-free galaxy \citep[e.g.][]{barnesFormationDwarfGalaxies1992, ducIdentificationOldTidal2014, ploeckingerTidalDwarfGalaxies2018}. Since the material used to build the galaxy has already been enriched in the more massive parent system(s), a metallicity higher than that predicted by the current-day mass-metallicity relation is expected -- as we observe for AndXIX\@. However, while AndXIX has a low dark matter \textit{density}, its total M/L ratio is high, indicating it is still a dark-matter-dominated system \citetalias{C19}. This is inconsistent with the lack of dark matter associated with TDGs. 

In addition, galaxies identified as TDGs in the literature are all typically young ($\lesssim4$~Gyr), gas-rich, and clearly located in the tidal tails of more massive systems. While AndXIX is in the vicinity of other stellar debris, as discussed in \S\ref{sec:tides}, there is currently no clear connection between this debris, AndXIX, and M31. More critically, AndXIX is both ancient \citepalias{Col22} and contains little to no gas \citep[e.g.][]{kaisinHaSurveyLowmass2013}, making it difficult to securely identify as a TDG. Proper motion measurements might help to establish whether AndXIX was ever located near a system sufficiently massive enough to form TDGs on a timescale consistent with that of its ancient star formation history. 

\section{Summary} \label{sec:concs}

In this work, we present the first $\alpha$-abundance measurements for the diffuse M31 satellite AndXIX, derived from medium-resolution spectra. From data first presented in \citetalias{C19}, we identify 86 targets which we consider to be likely AndXIX members, defined here as having $P_{\text{tot}}>0.5$ and $\Gamma/\sigma_\Gamma<0.2$. We group stars with similar photometric metallities and coadd them to obtain a total of 20 independent data points. We use spectral synthesis to derive [Fe/H] and [$\alpha$/Fe] measurements for each coadded spectrum; these reflect the weighted average of the underlying values of the individual contributing component stars. 

We find a trend of declining [$\alpha$/Fe] with increasing [Fe/H], as seen in other MW and M31 dSphs, indicating Type Ia SN contribute throughout the chemical evolution of AndXIX\@. The slope of this trend is steep (d[$\alpha$/Fe]/d[Fe/H]$\sim-0.65$), and comparable to that in other similar-luminosity dSphs also comprised primarily of ancient stellar populations (e.g., Sextans). The mean metallicity we measure for AndXIX ([Fe/H]$\sim1.5$) is $\sim$0.6~dex higher than that from existing literature, placing it $\sim2\sigma$ above the present-day stellar mass-metallicity relation for Local Group dSphs.

Our abundance measurements do not support mergers as a formation mechanism for AndXIX\@. Early high-speed, gas-rich mergers might explain a second burst of SF seen in AndXIX’s SFH, but the resulting vigorous star formation and increased rate of Type II SN should produce a shallow, if not flat slope in [$\alpha$/Fe]-[Fe/H] space. Late dry mergers of reionization fossils are also expected to result in a relatively flat [$\alpha$/Fe]-[Fe/H] slope but at very low metallicity, given the lack of time for significant numbers of Type Ia SN to occur in the component galaxies pre-reionization. The relatively high mean [Fe/H] and steep d[$\alpha$/Fe]/d[Fe/H] slope we measure for AndXIX cannot be easily explained in these scenarios. The TDG scenario for the formation of AndXIX could explain the higher than expected [Fe/H] we observe for it. However, the ancient age of AndXIX and the lack of clear connection to a larger parent system make this suggestion difficult to verify.  

Instead, we suggest tidal interactions as a promising avenue of formation for AndXIX\@. We posit a scenario where AndXIX was originally $\sim10\times$ more massive than its current luminosity would suggest, allowing it to retain gas and evolve to reach a higher [Fe/H] than satellites with a similar current luminosity. Since then, it has experienced a strong tidal interaction, stripping some of its existing stellar and dark mass: explaining the low DM halo density observed by \citetalias{C19}, and its diffuse structure. Either at the same time, or -- more likely -- beginning even earlier, AndXIX's gas has also been stripped, quenching it as indicated by its SFH (\citetalias{Col22}). This relatively ancient quenching prevented the extended star formation which results in the shallow d[$\alpha$/Fe]/d[Fe/H] gradients usually seen in more massive systems, explaining the steep decline we observe. Proper motion measurements that allow the orbit of AndXIX to be traced are necessary to confirm the timing and intensity of any such interaction, e.g., with M31. Nevertheless, our data clearly place stronger constraints on AndXIX's formation, providing new insight into this unique galaxy and those like it.

%% Also note that the acknowlodgment environment does not support long amounts of text. If you have a lot of people and institutions to acknowledge, do not use this command. Instead, create a new \section{Acknowledgments}.
\begin{acknowledgments}
	The authors recognize and acknowledge the very significant cultural role and reverence that the summit of Mauna Kea has always had within the indigenous Hawaiian community. We are most fortunate to have the opportunity to conduct observations from this mountain. This research is based on observations made with the NASA/ESA HST obtained from the Space Telescope Science Institute, which is operated by the Association of Universities for Research in Astronomy, Inc., under NASA contract NAS 5-26555. These observations are associated with program ID 15302; support for this work (L.R.C and K.M.G.) was provided by NASA through grant HST-GO-15302. I.E. acknowledges generous support from a Carnegie-Princeton Fellowship through Princeton University. %The analysis pipeline used to reduce the DEIMOS data was developed at UC Berkeley with support from NSF grant AST-0071048.
\end{acknowledgments}


\vspace{5mm}
\facilities{Keck:II(DEIMOS), Subaru(Suprime-Cam)}


\software{astropy \citep{astropycollaborationAstropyCommunityPython2013,astropycollaborationAstropyProjectBuilding2018a, astropycollaborationAstropyProjectSustaining2022}, corner \citep{foreman-mackeyCornerPyScatterplot2016}, dustmaps \citep{greenDustmapsPythonInterface2018}, emcee \citep{foreman-mackeyEmceeMCMCHammer2013}, matplotlib \citep{hunterMatplotlib2DGraphics2007}, numpy \citep{vanderwaltNumPyArrayStructure2011}, phew \citep{nunezPHEWPytHonEquivalent2022}, scipy \citep{virtanenSciPyFundamentalAlgorithms2020}} 


\appendix
\section{C19 member comparison}\label{sec:c19compapp}
Here, we discuss differences between our AndXIX sample and that of \citetalias{C19}. There are multiple factors which contribute to the differences in membership noted in \S\ref{sec:c19mcomp}: a) differing reduction methods for the data; b) different measurement techniques used to derive the velocities and/or the equivalent widths of the \ion{Na}{1} lines; and c) differing membership criteria used. We cannot disentangle the first two effects, which affect the entire underlying dataset; we therefore discuss these two effects first, followed by the effect of the differing membership criteria separately. 

\begin{figure*}
	\includegraphics[width=0.5\columnwidth]{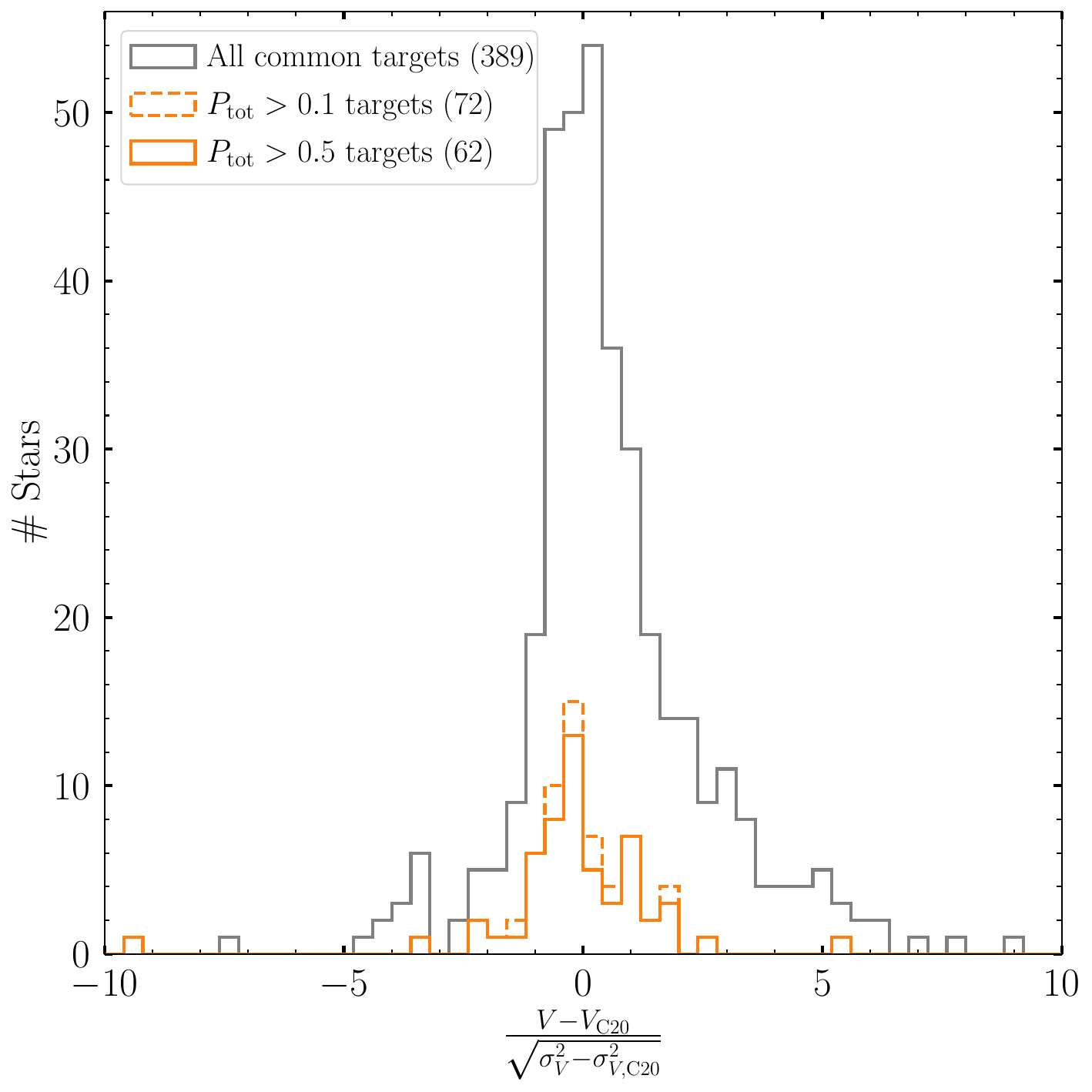}\includegraphics[width=0.5\columnwidth]{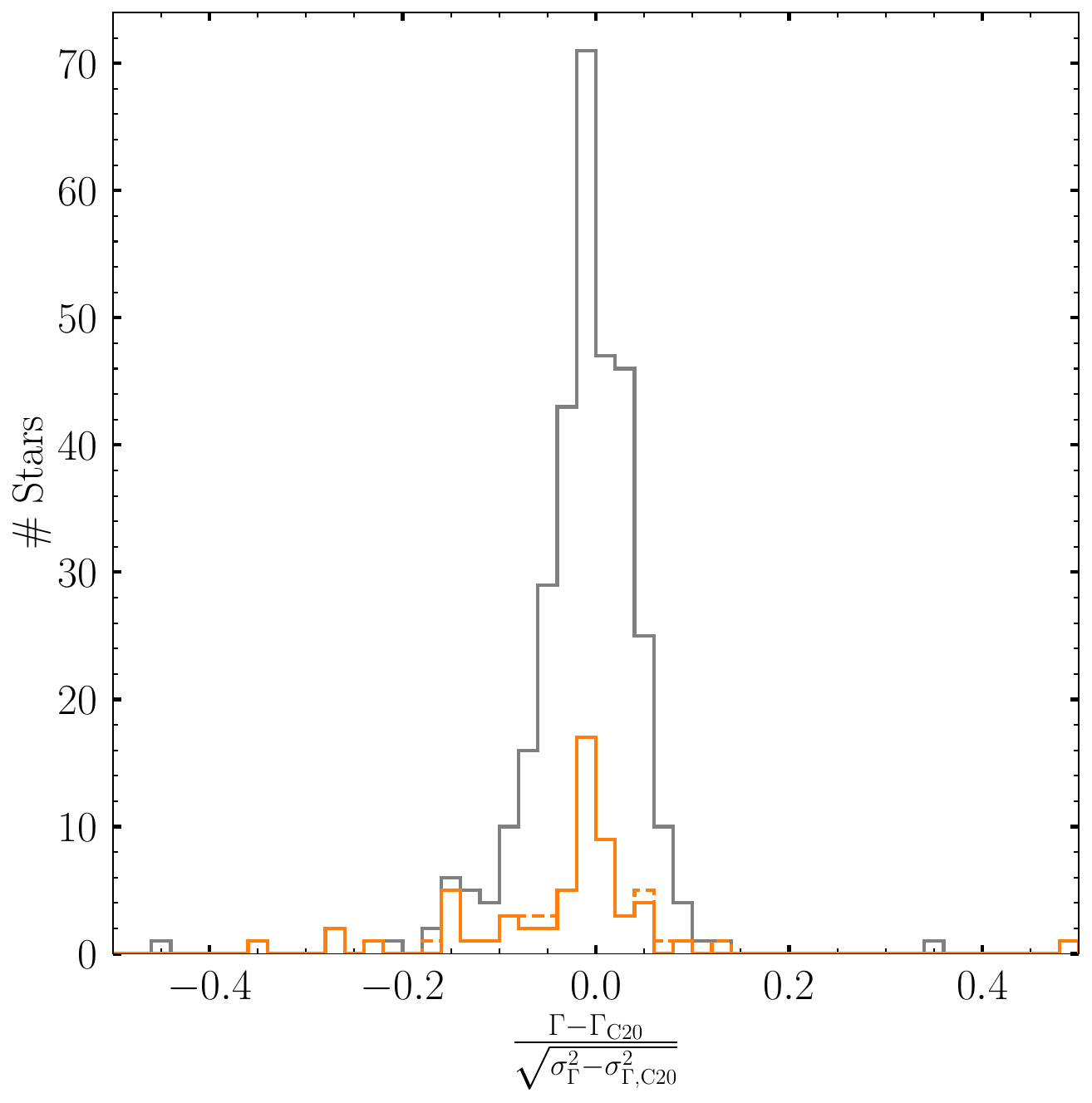}
	\caption{Normalised differences between the LOS velocity (left) and equivalent width of the \ion{Na}{1} doublet ($\Gamma$, right) produced from our reduction and that of \citetalias{C19}. The grey histogram presents the differences for the entire common sample of 389 targets; the dashed orange line indicates common targets which have a $P_{\text{tot}}>0.1$ of being associated with AndXIX (the cutoff used in \citetalias{C19}), and the solid orange line shows common targets which have a $P_{\text{tot}}>0.5$ of being associated with AndXIX. Stars which are likely associated with AndXIX have relatively small differences in their derived velocities and $\Gamma$ values derived from both reductions.}
	\label{fig:difhists}
\end{figure*}

There are no large systemic differences across the full underlying dataset: targets that are successfully reduced by both our pipeline and that of \citetalias{C19} have correlated LOS velocities and \ion{Na}{1} equivalent width measurements. We show difference histograms for the velocity and \ion{Na}{1} EWs, scaled by the uncertainty in the differences (i.e.\ $\frac{V-V_{\mathrm{C20}}}{\sqrt{\sigma_V^2-\sigma_{V,\mathrm{C20}}^2}}$ and $\frac{\Gamma-\Gamma_{\mathrm{C20}}}{\sqrt{\sigma_\Gamma^2-\sigma_{\Gamma,\mathrm{C20}}^2}}$) in Fig.~\ref{fig:difhists}. Grey lines indicate the full shared sample, while solid and dashed orange lines indicate targets which we calculate as having $P_{\text{tot}}>0.5$ and $P_{\text{tot}}>0.1$ (reflecting the different $P_{\text{tot}}$ cutoffs we use compared to \citetalias{C19}). We find that while there are certainly large tails to these distributions, particularly for likely member stars the median differences are consistent with zero. 

\begin{figure}
	\includegraphics[width=0.5\columnwidth]{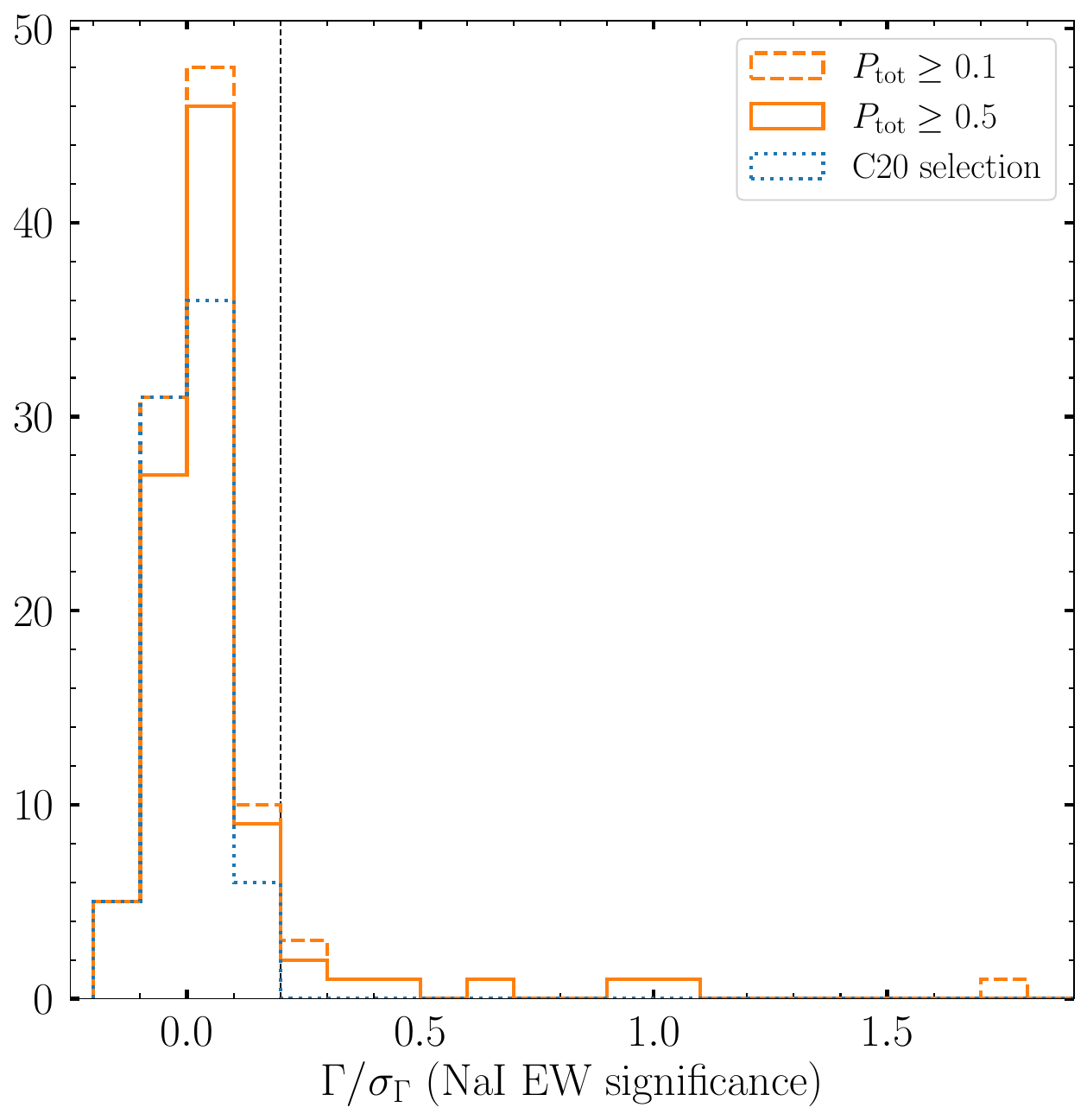}
	\caption{$\Gamma/\sigma_\Gamma$ histogram for targets with $P_{\text{tot}}>0.5$ (solid orange) and $P_{\text{tot}}>0.1$ (dashed orange). The black dashed line indicates a $\Gamma/\sigma_\Gamma$ cutoff of 0.2, which we use to select likely AndXIX members. The dashed blue histogram indicates targets instead selected using the criteria described in \protect\citetalias{C19}, i.e.\ $P_{\text{tot}}>0.1$ and $\Gamma<2$. All targets which pass the \citetalias{C19} selection also pass our $\Gamma/\sigma_\Gamma$ cut. }
	\label{fig:EWcomp}
\end{figure}

We now consider the effects of the different membership criteria we apply. Fig.~\ref{fig:EWcomp} presents a histogram of $\Gamma/\sigma_\Gamma$ for all targets with $P_{\text{tot}}>0.1$ (dashed orange) and $P_{\text{tot}}>0.5$ (solid orange), compared to that for stars which are selected according to the membership criteria of \citetalias{C19} (i.e.\ $P_{\text{tot}}>0.1$ and $\Gamma<2$) in blue. The vertical dashed black line indicates the $\Gamma/\sigma_\Gamma<0.2$ cutoff we apply to selected likely AndXIX members. %The right panel presents a $\Gamma$ histogram for the same subsamples. 
We find that the two membership criteria produce very similar samples, with 85\% of likely members in common between the two cuts. The higher $P_{\text{tot}}$ cutoff for our primary subsample reduces the number of likely AndXIX stars by 7; as discussed in \S\ref{sec:c19mcomp}, we coadd these stars separately as our “low-probability” subsample. All stars which have $\Gamma<2$ pass our $\Gamma/\sigma_\Gamma$ cut; but our $\Gamma/\sigma_\Gamma$ cut does result in an additional 16 stars being considered members. These stars have very low S/N, which makes any nominal measurement of their \ion{Na}{1} absorption very uncertain. We confirm that none of these stars have visually identifiable \ion{Na}{1} doublets. 

\bibliographystyle{aasjournal}
\bibliography{andxix.bib,software.bib}



\end{document}